\documentclass[journal]{IEEEtran}
\usepackage{cite}
\usepackage{amsmath,amssymb,amsfonts}
\usepackage{algorithmic}
\usepackage{graphicx}
\usepackage{textcomp}

\newtheorem{thm}{Theorem}

\newtheorem{rem}[thm]{Remark}

\usepackage{times}
\usepackage{algorithm}
\usepackage{algorithmic}
\usepackage{varwidth}
\usepackage{caption}

\begin{document}
\title{A Levenberg-Marquardt algorithm for sparse identification of dynamical systems}
\author{Mark Haring, Esten Ingar Gr\o tli, Signe Riemer-S\o rensen, Katrine Seel, and Kristian Gaustad Hanssen
\thanks{This work was supported by the industry partners Borregaard, Elkem, Hydro, Yara and the Research Council of Norway through the project TAPI: Towards Autonomy in Process Industries, project number 294544.}
\thanks{Mark Haring, Esten Ingar Gr\o tli, Signe Riemer-S\o rensen, and Kristian Gaustad Hanssen are with the Department of Mathematics and Cybernetics, SINTEF Digital, Trondheim/Oslo, Norway (email: \mbox{mark.haring@sintef.no}, \mbox{esteningar.grotli@sintef.no}, \mbox{signe.riemer-sorensen@sintef.no}, \mbox{kristian.gaustad.hanssen@sintef.no}).}
\thanks{Katrine Seel is with the Department of Engineering Cybernetics, Norwegian University of Science and Technology (NTNU), 7491 Trondheim, Norway, and with the Department of Mathematics and Cybernetics, SINTEF Digital, Trondheim/Oslo, Norway (email: \mbox{katrine.seel@sintef.no}).}}

%



\maketitle

\begin{abstract}
Low complexity of a system model is essential for its use in real-time applications. However, sparse identification methods commonly have stringent requirements that exclude them from being applied in an industrial setting. In this paper, we introduce a flexible method for the sparse identification of dynamical systems described by ordinary differential equations. Our method relieves many of the requirements imposed by other methods that relate to the structure of the model and the data set, such as fixed sampling rates, full state measurements, and linearity of the model. The Levenberg-Marquardt algorithm is used to solve the identification problem. We show that the Levenberg-Marquardt algorithm can be written in a form that enables parallel computing, which greatly diminishes the time required to solve the identification problem. An efficient backward elimination strategy is presented to construct a lean system model.
\end{abstract}


\IEEEpeerreviewmaketitle

\section{Introduction}
\label{sec:introduction}

\IEEEPARstart{I}{n} many practical applications, a mathematical model of a dynamical system is a prerequisite to understanding, predicting and manipulating the behavior of the system in an effective manner. Obtaining a suitable model can be a challenging task. For some systems, first principles (e.g., established laws of physics) may be used to derive model equations that represent the dominant system dynamics. However, for many other systems, this way of model discovery is out of the question due to insufficient information about the system. Moreover, even if first principles can be used to obtain a model, the resulting model may be too complex to use in real-time applications for design, optimization and control. For example, in areas such as fluid dynamics, electrodynamics, and quantum mechanics, a first-principle model may consist of multiple partial differential equations for which the evolution of the state of the system is hard to compute, let alone in real time.

Alternatively, we may construct a system model using data following a system identification or machine learning approach. The corresponding system model can often be written in the form of an artificial neural network, where the type of neurons can range from commonly used neurons with sigmoidal or rectified linear activation functions to wavelets, splines, radial basis functions, monomials or Gaussian processes, to name a few \cite{Sjoberg1995}. Generally designed to be a universal function approximator, the use of an artificial neural network often leads to an unnecessarily complex model. High complexity strongly hampers the use of the model in real-time applications, and it commonly goes hand in hand with stringent data requirements. It should be noted that the complexity of the model strongly depends on the chosen coordinate frame and type of neurons. As a first example, the orbits of the planets in our solar system can be described much simpler and in fewer terms by taking the Sun as center of reference than Earth. As mentioned in \cite{Brunton2016} and \cite{Rudy2017}, many physical systems allow for a sparse representation in a suitable coordinate frame. As a second example, it takes a large artificial network with rectified linear units to accurately approximate a quadratic function on a large (but finite) domain, while it requires only three monomials to do the same (i.e., monomials of degrees zero, one, and two, respectively). Training such a large artificial network needs many measurements of the quadratic function, while fitting the three monomials requires a minimum of only three data points (disregarding measurement noise). Without any knowledge of the system to go by, the chance that a chosen coordinate frame and neuron type result in an accurate model of low complexity is generally slim.

In many industrial applications, neither a first-principle model nor a purely data-driven model is constructed with high accuracy due to various reasons including unknown environmental conditions, a varying composition of raw materials, knowledge caveats, imprecise data registration, and a lack of exploration. In addition, there may be parts of the system for which only a few measurements are available due to a high cost or inability to conduct more. What complicates matters further is that these measurements may have irregular sampling rates because smart sampling algorithms (see \cite{Nduhura-Munga2013}) have been applied or measurements have been conducted manually. To compose accurate models, one is often forced to combine both first-principle information and measurement data. These resulting grey-box models come in many different forms \cite{Ljung2010}. In an attempt to improve the model accuracy while keeping the model complexity low, we can use an artificial neural network to model only those parts of the system for which no first-principle model is available or for which the solution of the first-principle model is difficult to compute or inaccurate. However, due to a high network complexity, this alone is generally not enough to ensure that the resulting model is simple enough to be used for real-time applications. 

In this paper, we apply sparse identification to derive a low-complexity network model. The paper is inspired by the works in \cite{Brunton2016} and \cite{Rudy2017}. In these, a lean feedforward network is obtained by removing all edges (i.e., the connections between the neurons in the network) that have little or no influence on the quality of the fit. The network can be pruned by adding a sparsity-promoting regularization term to the cost function (or loss function) that is used to train the network; see lasso \cite{Tibshirani1996} and sparse relaxed regularized regression \cite{Zheng2019}, for instance. Alternatively, a sequential algorithm, such as stepwise regression \cite{Efroymson1960} or sequential thresholding \cite{Brunton2016}, may be applied to remove all irrelevant edges in an iterative fashion. Other model-selection approaches are discussed in \cite{Hong2008} and \cite{Miller2002}. The sparse-identification methods in \cite{Brunton2016} and \cite{Rudy2017} rely on linear regression. Although they are relatively simple and fast enough to perform model identification in real time \cite{Kaiser2018,Quade2018}, they rely on the limiting assumptions that the network model is linear in all its parameters, and that all system states are measured and their first-order time derivatives can be accurately approximated. This last condition commonly translates to the requirement of a sufficiently high sampling rate for all measurements in order to mitigate the negative influences of measurement noise by averaging over multiple samples. In \cite{Mauroy2019}, the computation of time derivatives of the states is based on Koopman theory. This technique can be applied to low-rate sampled data, but the requirement of a fixed sampling rate is often too restrictive to be applied to industrial data.

If the model is composed of differential equations, we may apply numerical integration to compute the state values of the system at any point in time and compare these to the measurements to optimize the model. This removes the requirement that the network should be linear in its parameters. Moreover, an arbitrarily sampled data set can be used. In \cite{Rudy2019}, the solutions of differential equations modeled by deep neural networks are computed using a Runge-Kutta scheme. In \cite{Raissi2018b}, the numerical integration algorithm is directly encoded in the kernels of the Gaussian processes. As an alternative to traditional numerical integration, a neural network can be trained to produce solutions of differential equations \cite{Lagaris1998,Li2020}. The deep neural network in \cite{Raissi2019} is designed to obtain solutions of differential equations and identify model parameters. To the best of our knowledge, no attempts of network sparsification have been made for any of the identification methods that rely on numerical integration. 

A possible reason for this is the high computational demand for training the network. Numerical integration of the state equations implies calculating the state values on a finite (and flexible) time grid. Evaluating the cost function used for training the network subsequently requires computing the state values for all grid points. The effective dimension of the optimization problem to be solved for training the network is equal to the number of network parameters plus the number of state variables times the number of grid points. A large number of grid points may be required to accurately approximate the state solution, especially if the time series of measurements span a large time interval. Therefore, the dimension of the optimization problem may be very large. In turn, the optimization problem may be difficult to solve. Note that network pruning would require solving an even harder regularized optimization problem or retraining the network multiple times using a backward elimination strategy.

In this paper, we develop an efficient method for training an artificial neural network  for system models described by ordinary differential equations. Common training algorithms include gradient descent, the limited-memory Broyden–Fletcher–Goldfarb–Shanno (L-BFGS) algorithm and the Levenberg-Marquardt algorithm \cite{Le2011,Mukherjee2012,Tan2019}. Compared to the first-order convergence of gradient descent, a faster second-order convergence can be obtained using L-BFGS or Levenberg-Marquardt. The Levenberg-Marquardt algorithm has the advantage that it is relatively easy to parallelize. Due to the Markovian nature of the system model, we can rewrite the approximation of the optimization problem by Levenberg and Marquardt as batches of recursive small-scale optimization subproblems. We effectively divide the time interval from the first to the last measurement in separate subintervals, where each subinterval corresponds to one batch of optimization subproblems. All batches can be solved in parallel. Subsequently, the solution to the original approximation can be reconstructed by combining the results from all batches. Parallelization can significantly reduce the wall time required to train the network. In addition, we outline how a similar problem formulation can be used to prune the network using backward elimination by removing irrelevant network edges one at a time.

The paper is organized as follows. A formulation of the system identification problem is given in Section~\ref{sec:problem_formulation}. The standard approach of solving the system identification problem using the Levenberg-Marquardt algorithm is discussed in Section~\ref{sec:solving_problem}. In Section~\ref{sec:exploiting_structure}, it is shown how the solution method can be rewritten in order to parallelize parts of the algorithm. In Section~\ref{sec:network_spasification}, a method for network sparsification using backward elimination is presented that utilizes a similar approach as outlined in Section~\ref{sec:exploiting_structure} to efficiently approximate the effect of removing a network edge on the overall quality of the fit. In Section~\ref{sec:examples}, the sparsification method is illustrated by means of two examples. Advantages and drawbacks of the proposed methodology are summarized in Section~\ref{sec:discussion}.

\subsection{Notations}

The sets of real numbers, nonnegative real numbers and positive real numbers are denoted by $\mathbb{R}$, $\mathbb{R}_{\geq 0}$  and $\mathbb{R}_{>0}$, respectively. $\mathbb{N}$ and $\mathbb{N}_{>0}$ denote the sets of natural numbers (i.e. nonnegative integers) and positive integers. The ceiling function that maps a real number $\alpha$ to the least integer greater than or equal to $\alpha$ is denoted by $\lceil\alpha\rceil$. The transpose of any vector $\mathbf{x}$ is written as $\mathbf{x}^T$. For any nonsingular, square matrix $\mathbf{A}$, its inverse is denoted by $\mathbf{A}^{-1}$. Similarly, $\mathbf{A}^{-T}$ is the transpose of the inverse. The Moore-Penrose pseudoinverse of the matrix $\mathbf{M}$ is written as $\mathbf{M}^+$. We denote the union of two sets $\mathcal{B}$ and $\mathcal{C}$ by $\mathcal{B} \cup \mathcal{C}$. If $\mathcal{C}$ is a subset of $\mathcal{B}$, then the set difference of $\mathcal{B}$ and $\mathcal{C}$ (i.e. the relative complement of $\mathcal{C}$ in $\mathcal{B}$) is written as $\mathcal{B} \setminus \mathcal{C}$. Let $\mathcal{T} = \{t_1, \, t_2, \, \dots, \, t_n\}$ be an ordered set with $n$ elements. The notation $\{\mathbf{y}(t)\}_{t \in \mathcal{T}}$ is short for $\{\mathbf{y}(t_1), \, \mathbf{y}(t_2), \, \dots, \, \mathbf{y}(t_n)\}$. For any time-varying vector signal $\mathbf{s}(t)$, where $t$ denotes the time, we write $\frac{d \mathbf{s}}{dt}(t)$ as $\mathbf{\dot{s}}(t)$. The Euclidean norm is denoted by $\| \cdot \|$.

\section{Formulation of the system identification problem}
\label{sec:problem_formulation}

Let the state trajectories of a dynamical system be given by the solutions of the following ordinary-differential equations that comprise a first-principle model:
\begin{equation} \label{eq:dynamic_grey-box_model}
\mathbf{\dot{x}}(t) = \mathbf{f}_{\text{phys}}(t,\mathbf{x}(t)) + \mathbf{e}(t).
\end{equation}
Here, $\mathbf{x}(t) \in \mathbb{R}^{n_\mathbf{x}}$ is the state vector with dimension $n_\mathbf{x} \in \mathbb{N}_{>0}$, $\mathbf{f}_{\text{phys}}: \mathbb{R} \times \mathbb{R}^{n_\mathbf{x}} \rightarrow \mathbb{R}^{n_\mathbf{x}}$ is a continuously differentiable function, $\mathbf{e}(t) \in \mathbb{R}^{n_\mathbf{x}}$ is the model error, and $t \in \mathbb{R}$ denotes the time. The exact value of the state is unknown in most practical scenarios. To improve the accuracy of the first-principle model, we approximate the error by a continuously differentiable feedforward network $\mathbf{f}_{\text{net}}: \mathbb{R}^{n_\mathbf{x}} \times \mathbb{R}^{n_\mathbf{u}} \times \mathbb{R}^{n_\mathbf{a}} \rightarrow \mathbb{R}^{n_\mathbf{x}}$. That is,
\begin{equation} \label{eq:error_model}
\mathbf{e}(t) \approx \mathbf{f}_{\text{net}}(\mathbf{x}(t),\mathbf{u}(t),\mathbf{a}).
\end{equation}
The network consists of an input layer, one or multiple hidden layers, and an output layer. The inputs of the network are the state $\mathbf{x}(t)$ and the vector of known time-varying functions $\mathbf{u}(t) \in \mathbb{R}^{n_\mathbf{u}}$ with dimension $n_\mathbf{u} \in \mathbb{N}$. The output of the network is the estimate of the error in the first-principle model. The vector of network parameters (e.g., weights and biases) is given by $\mathbf{a} \in \mathbb{R}^{n_\mathbf{a}}$, where $n_\mathbf{a} \in \mathbb{N}_{>0}$ is its dimension. We do not pose any restrictions on the structure of the network or the choice of neurons. Note that this formulation can be easily adapted to include other kinds of error models, such as the closure models in \cite{Pan2018}. Although the focus of this paper is to improve the accuracy of the first-principle model by approximating the model error with a feedforward network, we note that this formulation can be made more general by considering models of the form:
\begin{equation} \label{eq:continuous_state_model}
\mathbf{\dot{x}}(t) \approx \mathbf{f}(t,\mathbf{x}(t),\mathbf{a}),
\end{equation}
where $\mathbf{a}$ is allowed to be any parameter vector. The only essential requirement is that the function $\mathbf{f}: \mathbb{R} \times \mathbb{R}^{n_\mathbf{x}} \times \mathbb{R}^{n_\mathbf{a}} \rightarrow \mathbb{R}^{n_\mathbf{x}}$ is differentiable with respect to the state and parameter vectors. Obviously, the system model consisting of the first-principle model in \eqref{eq:dynamic_grey-box_model} and the error model in \eqref{eq:error_model} can be written in this form.

To identify suitable parameter values, we compare the state solutions of the model with process measurements. Consider a finite time series of process measurements. Let $\mathcal{T}_m \subset \mathbb{R}$ be a finite set containing all measurement times.  We denote by $\mathbf{y}(t_m) \in \mathbb{R}^{n_\mathbf{y}}$ the vector of measurements with dimension $n_\mathbf{y} \in \mathbb{N}_{>0}$ taken at time $t = t_m \in \mathcal{T}_m$. The dimension $n_\mathbf{y}$ is allowed to be time varying (i.e., $n_\mathbf{y} = n_\mathbf{y}(t_m)$). The relation between the measurements and the state is modeled by a continuously differentiable function $\mathbf{h}: \mathcal{T}_m \times \mathbb{R}^{n_\mathbf{x}} \rightarrow \mathbb{R}^{n_\mathbf{y}}$, such that
\begin{equation} \label{eq:discrete_measurement_model}
\mathbf{y}(t_m) \approx \mathbf{h}(t_m, \mathbf{x}(t_m))
\end{equation}
for all $t = t_m \in \mathcal{T}_m$. Note that the sampling of the measurements can be arbitrary. Notation-wise, it is convenient to number the measurements. We define
\begin{equation}
\mathcal{I}_m = \left\{i \in\mathbb{N}: \, 1 \leq i \leq N_m \right\},
\end{equation}
where $N_m \in \mathbb{N}_{>0}$ is the cardinality of $\mathcal{T}_m$ (i.e., the set $\mathcal{T}_m$ contains $N_m$ elements). For any $i \in \mathcal{I}_m$, $t_m^{(i)}$ denotes the $i^\text{th}$ element of the set $\mathcal{T}_m$ in ascending order. In turn, the $i^\text{th}$ vector of measurements is denoted by $\mathbf{y}(t_m^{(i)})$. 

Obtaining the parameter values that fit the measurements best requires that the state and the parameters of the model are identified simultaneously. To find suitable values, 
we compute the minimizer of the nonlinear least-squares problem:
\begin{equation} \label{eq:least-squares_continuous_discrete}
\begin{aligned}
&\min_{ \mathbf{a}, \, \{\mathbf{x}\}, \, \{\boldsymbol\varepsilon_\mathbf{x}\}, \, \{\boldsymbol\varepsilon_\mathbf{y}\} } \Biggl\{ \int_{\underline{t}_m}^{\overline{t}_m}  \boldsymbol\varepsilon_\mathbf{x}^T(t) \mathbf{W}_\mathbf{x}^{-1}(t) \boldsymbol\varepsilon_\mathbf{x}(t)  dt + \mu_\mathbf{a} \|  \mathbf{a} \|^2 \\
& \quad + \sum_{t_m \in \mathcal{T}_m} \boldsymbol\varepsilon_\mathbf{y}^T(t_m) \mathbf{W}_\mathbf{y}^{-1}(t_m) \boldsymbol\varepsilon_\mathbf{y}(t_m)   + \mu_\mathbf{x} \int_{\underline{t}_m}^{\overline{t}_m} \| \mathbf{x}(t) \|^2 dt:\\
& \qquad  \left(\forall t \in [\underline{t}_m, \, \overline{t}_m] \right) \left[ \mathbf{\dot{x}}(t) = \mathbf{f}(t,\mathbf{x}(t),\mathbf{a}) + \boldsymbol\varepsilon_\mathbf{x}(t) \right], \,\\
& \qquad \left(\forall t_m \in \mathcal{T}_m \right) \left[ \mathbf{y}(t_m) = \mathbf{h}(t_m,\mathbf{x}(t_m)) + \boldsymbol\varepsilon_\mathbf{y}(t_m) \right]
\Bigg\}
\end{aligned}
\end{equation}
where $\{\mathbf{x}\}$ and $\{\boldsymbol\varepsilon_\mathbf{x}\}$ are shorthand notations for the state $\{\mathbf{x}(t)\}_{t = \underline{t}_m}^{\overline{t}_m}$ and the state error $\{\boldsymbol\varepsilon_\mathbf{x}(t)\}_{t = \underline{t}_m}^{\overline{t}_m}$, respectively, and where $\{\boldsymbol\varepsilon_\mathbf{y}\}$ is short for the set $\{\boldsymbol\varepsilon_\mathbf{y}(t_m) \}_{t_m \in \mathcal{T}_m}$, with modeling error $\boldsymbol\varepsilon_\mathbf{y}(t_m) \in \mathbb{R}^{n_\mathbf{y}}$ for all $t_m \in \mathcal{T}_m$. Here, $\underline{t}_m = t_m^{(1)}$ and $\overline{t}_m = t_m^{(N_m)}$ are the times of the first and last measurement, respectively. The symmetric, positive-definite matrices $\mathbf{W}_\mathbf{x}(t) \in \mathbb{R}^{n_\mathbf{x} \times n_\mathbf{x}}$ and $\mathbf{W}_\mathbf{y}(t_m) \in \mathbb{R}^{n_\mathbf{y} \times n_\mathbf{y}}$ are weighting matrices of the model errors $\boldsymbol\varepsilon_\mathbf{x}(t)$ and $\boldsymbol\varepsilon_\mathbf{y}(t_m)$, respectively. If $\boldsymbol\varepsilon_\mathbf{x}(t)$ and $\boldsymbol\varepsilon_\mathbf{y}(t_m)$ are regarded as vectors of stochastic variables, then $\mathbf{W}_\mathbf{x}(t)$ and $\mathbf{W}_\mathbf{y}(t_m)$ can be chosen equal to the covariance matrix of $\boldsymbol\varepsilon_\mathbf{x}(t)$ and $\boldsymbol\varepsilon_\mathbf{y}(t_m)$ in order to obtain an estimate with minimal variance. A more pragmatic approach is to think of $\mathbf{W}_\mathbf{x}(t)$ and $\mathbf{W}_\mathbf{y}(t_m)$ as scaling matrices, such that $\mathbf{S}_\mathbf{x}^{-1}(t) \boldsymbol\varepsilon_\mathbf{x}(t)$ and $\mathbf{S}_\mathbf{y}^{-T}(t_m) \boldsymbol\varepsilon_\mathbf{y}(t_m)$ are vectors of elements with roughly equal magnitude based on prior knowledge, where
\begin{equation} \label{eq:S_x}
\mathbf{W}_\mathbf{x}(t) = \mathbf{S}_\mathbf{x}(t) \mathbf{S}_\mathbf{x}^T(t)
\end{equation}
and
\begin{equation} \label{eq:S_y}
\mathbf{W}_\mathbf{y}(t_m) = \mathbf{S}_\mathbf{y}(t_m) \mathbf{S}_\mathbf{y}^T(t_m).
\end{equation}
The terms in the cost function related to the (small) constants $\mu_\mathbf{x}, \mu_\mathbf{a} \in \mathbb{R}_{\geq 0}$ are regularization terms to remedy the effects of overfitting \cite{Ying2019}. If multiple or even infinitely many minima exist, these regularization terms keep the magnitudes of the computed, minimizing state and parameter values relatively small. The selection of suitable regularization constants is an open problem, and a topic for future work.

\section{Solving the system identification problem using the Levenberg-Marquardt algorithm}
\label{sec:solving_problem}

\begin{figure*}[!t]
\addtocounter{equation}{3}
\vspace*{-4pt}
\normalsize
\begin{equation} \label{eq:least-squares_discretized}
\begin{aligned}
&\min_{ \mathbf{a}, \, \{\mathbf{x}\}, \, \{\boldsymbol\varepsilon_\mathbf{x}\}, \, \{\boldsymbol\varepsilon_\mathbf{y}\} } \Biggl\{ \sum_{j \in \mathcal{J}_d\setminus\{N_d\}}  \boldsymbol\varepsilon_\mathbf{x}^T(t_d^{(j{+}\frac{1}{2})}) \mathbf{W}_\mathbf{x}^{-1}(t_d^{(j{+}\frac{1}{2})}) \boldsymbol\varepsilon_\mathbf{x}(t_d^{(j{+}\frac{1}{2})})  \Delta t_d^{(j{+}\frac{1}{2})} + \sum_{j_m \in \mathcal{J}_m} \boldsymbol\varepsilon_\mathbf{y}^T(t_d^{(j_m)}) \mathbf{W}_\mathbf{y}^{-1}(t_d^{(j_m)}) \boldsymbol\varepsilon_\mathbf{y}(t_d^{(j_m)}) + \mu_\mathbf{a} \| \mathbf{a} \|^2 \\
& \quad + \frac{\mu_\mathbf{x}}{2}\sum_{j \in \mathcal{J}_d\setminus\{N_d\}} \left( \| \mathbf{x}(t_d^{(j)}) \|^2 + \| \mathbf{x}(t_d^{(j{+}1)}) \|^2  \right) \Delta t_d^{(j{+}\frac{1}{2})}  : \,  \left(\forall j_m \in \mathcal{J}_m \right) \left[ \mathbf{y}(t_d^{(j_m)}) = \mathbf{h}(t_d^{(j_m)},\mathbf{x}(t_d^{(j_m)})) + \boldsymbol\varepsilon_\mathbf{y}(t_d^{(j_m)}) \right], \\
& \quad \left(\forall j \in \mathcal{J}_d\setminus\{N_d\} \right) \left[ \frac{\mathbf{x}(t_d^{(j{+}1)}) - \mathbf{x}(t_d^{(j)})}{ \Delta t_d^{(j{+}\frac{1}{2})} } = \mathbf{f}(t_d^{(j{+}\frac{1}{2})}, {\textstyle\frac{1}{2}} ( \mathbf{x}(t_d^{(j)}) + \mathbf{x}(t_d^{(j{+}1)}) ),\mathbf{a}) + \boldsymbol\varepsilon_\mathbf{x}(t_d^{(j{+}\frac{1}{2})}) \right]
\Bigg\}
\end{aligned}
\end{equation}
\hrulefill
\vspace*{4pt}
\addtocounter{equation}{-4}
\end{figure*}
In order to compute the solution of the optimization problem in \eqref{eq:least-squares_continuous_discrete}, we first discretize the problem in time. Let $\Delta t \in \mathbb{R}_{>0}$ be the maximal step size of the discretization. Note that all continuous-time signals in \eqref{eq:least-squares_continuous_discrete} are defined on the interval $[\underline{t}_m, \, \overline{t}_m]$. Let $\mathcal{T}_d \subset \mathbb{R}$ be a set of $N_d \in \mathbb{N}_{>0}$ discretization points that are chosen such that $\mathcal{T}_m \subseteq \mathcal{T}_d$, and the distance between all subsequent points is smaller than or equal to $\Delta t$. A routine to generate such set is outlined in Appendix~\ref{app:discretization_time}. To make the proceeding notations easier, we define the index set
\begin{equation} \label{eq:definition_Jd}
\mathcal{J}_d= \left\{j \in\mathbb{N}: \, 1 \leq j \leq N_d \right\}.
\end{equation}
For any $j \in \mathcal{J}_d$, we denote by $t_d^{(j)}$ the $j^\text{th}$ element of $\mathcal{T}_d$ in ascending order. In addition, we define the index set
\begin{equation} \label{eq:definition_Jm}
\mathcal{J}_m = \left\{j \in \mathcal{J}_d: \, (\exists t_m \in \mathcal{T}_m)\left[t_d^{(j)} = t_m \right] \right\}
\end{equation}
such that $t_d^{(j_m)} \in \mathcal{T}_m$ for all $j_m \in \mathcal{J}_m$. The (variable) discretization step size is given by 
\begin{equation}
\Delta t_d^{(j{+}\frac{1}{2})} = t_d^{(j{+}1)} - t_d^{(j)}
\end{equation}
for all $j \in \mathcal{J}_d\setminus\{N_d\}$. 

\addtocounter{equation}{1}
We discretize the least-squares problem in \eqref{eq:least-squares_continuous_discrete} using the midpoint rule and the trapezoidal rule. With a minor abuse of notation, we get the optimization problem in \eqref{eq:least-squares_discretized}, with
\begin{equation}
t_d^{(j{+}\frac{1}{2})} = \frac{1}{2} \left( t_d^{(j{+}1)} + t_d^{(j)} \right)
\end{equation}
for all $j \in \mathcal{J}_d\setminus\{N_d\}$.
The sets $\{\mathbf{x}\}$, $\{\boldsymbol\varepsilon_\mathbf{x}\}$ and $\{\boldsymbol\varepsilon_\mathbf{y}\}$ are now short for $\{\mathbf{x}(t_d^{(j)})\}_{j \in \mathcal{J}_d}$, $\{\boldsymbol\varepsilon_\mathbf{x}(t_d^{(j)})\}_{j \in \mathcal{J}_d}$ and  $\{\boldsymbol\varepsilon_\mathbf{y}(t_d^{(j_m)})\}_{j_m \in \mathcal{J}_m}$, respectively. Because the discretization errors of the midpoint rule and the trapezoidal rule for each discretization step are $\mathcal{O}(\Delta t^3)$ and the total number of discretization steps is $\mathcal{O}(\Delta t^{-1})$, the corresponding difference between the solutions of \eqref{eq:least-squares_continuous_discrete} and \eqref{eq:least-squares_discretized} for all discretization steps combined is $\mathcal{O}(\Delta t^2)$. Hence, the solutions are identical in the limit as $\Delta t$ approaches zero. 

\begin{rem}
The midpoint rule and the trapezoidal rule are not the only numerical integration methods that can be applied to discretize the optimization problem in \eqref{eq:least-squares_continuous_discrete}. However, any such integration method is required to be explicit in order to apply the Levenberg-Marquardt algorithm later in this section. Alternatively, we may change the problem formulation in \eqref{eq:least-squares_continuous_discrete} by replacing all integrals by finite, weighted sums of function evaluations, as in \cite{Rudy2019}, for instance. However, the solution of this altered optimization problem may not correspond to the solution of \eqref{eq:least-squares_continuous_discrete}, not even in the limit.
\end{rem}

We may simplify the optimization problem in \eqref{eq:least-squares_discretized} by eliminating the variables of the model errors  $\{\boldsymbol\varepsilon_\mathbf{x}\}$ and $\{\boldsymbol\varepsilon_\mathbf{y}\}$  using the constraint equations. Subsequently, the optimization problem can be written as
\begin{equation} \label{eq:least-squares_problem_continuous_rewritten}
\min_{\mathbf{b}} \left\| \mathbf{g}(\mathbf{b}) \right\|^2,
\end{equation}
with
\begin{equation} \label{eq:definition_b}
\mathbf{b} = \begin{bmatrix}
 \mathbf{x}^T(t_d^{(1)}), \, \mathbf{x}^T(t_d^{(2)}), \, \dots, \, \mathbf{x}^T(t_d^{(N_d)}), \, \mathbf{a}^T
\end{bmatrix}^T
\end{equation}
and
\begin{equation} \label{eq:function_g_b}
\mathbf{g}(\mathbf{b}) =
\begin{bmatrix}
\mathbf{p}_{\mathbf{x},1}(\mathbf{x}(t_d^{(1)}),\mathbf{x}(t_d^{(2)}),\mathbf{a}) \\
\mathbf{p}_{\mathbf{x},2}(\mathbf{x}(t_d^{(2)}),\mathbf{x}(t_d^{(3)}),\mathbf{a}) \\
\vdots \\
\mathbf{p}_{\mathbf{x},N_d{-}1}(\mathbf{x}(t_d^{(N_d{-}1)}),\mathbf{x}(t_d^{(N_d)}),\mathbf{a}) \\
\mathbf{p}_{\mathbf{x},N_d}(\mathbf{x}(t_d^{(N_d)}))\\
\mathbf{p}_{\mathbf{a}}(\mathbf{a})
\end{bmatrix},
\end{equation}
where
\begin{equation}
\begin{aligned}
&\mathbf{p}_{\mathbf{x},j}(\mathbf{x},\mathbf{z},\mathbf{a}) \\
&=
\begin{bmatrix}
\scriptstyle{\mathbf{S}_\mathbf{x}^{-1}(t_d^{(j{+}\frac{1}{2})}) \left( \frac{\scriptstyle{\mathbf{z} - \mathbf{x}}}{\Delta t_d^{(j{+}\frac{1}{2})}} - \mathbf{f}(t_d^{(j{+}\frac{1}{2})},{\textstyle\frac{1}{2}}\left(\mathbf{x} + \mathbf{z} \right),\mathbf{a}) \right)  \sqrt{\Delta t_d^{(j{+}\frac{1}{2})}} }\\
\sqrt{\frac{\mu_\mathbf{x}}{2}} \mathbf{x}  \sqrt{\Delta t_d^{(j{+}\frac{1}{2})}} \\
\sqrt{\frac{\mu_\mathbf{x}}{2}} \mathbf{z} \sqrt{\Delta t_d^{(j{+}\frac{1}{2})}} \\
\left\{\begin{aligned}
& \mathbf{S}_\mathbf{y}^{-1}(t_d^{(j)})  \left( \mathbf{y}(t_d^{(j)}) - \mathbf{h}(t_d^{(j)},\mathbf{x}) \right), & &\mbox{if $j \in \mathcal{J}_m$}\\
& 0, & &\mbox{if $j \not\in \mathcal{J}_m$}
\end{aligned} \right.
\end{bmatrix} 
\end{aligned}
\end{equation}
for all $j \in \mathcal{J}_d\setminus\{N_d\}$,
\begin{equation}
\mathbf{p}_{\mathbf{x},N_d}(\mathbf{x}) = \mathbf{S}_\mathbf{y}^{-1}(t_d^{(N_d)})  \left( \mathbf{y}(t_d^{(N_d)}) - \mathbf{h}(t_d^{(N_d)},\mathbf{x}) \right),
\end{equation}
and
\begin{equation}
\mathbf{p}_{\mathbf{a}}(\mathbf{a}) = \sqrt{\mu_\mathbf{a}} \mathbf{a}.
\end{equation}
The optimization problem in \eqref{eq:least-squares_problem_continuous_rewritten} may be solved iteratively. Let $k \in \mathbb{N}$ be the iteration number. Writing the values of the optimization variables in \eqref{eq:definition_b} at iteration $k$ as $\mathbf{b}_k$, the first-order Taylor series approximation
\begin{equation} \label{eq:function_g_approximation}
\mathbf{g}(\mathbf{b}_{k+1}) \approx \mathbf{g}(\mathbf{b}_k) + \frac{d \mathbf{g}}{d \mathbf{b}}(\mathbf{b}_k) \boldsymbol\gamma_k
\end{equation}
is accurate if $\boldsymbol\gamma_k$ is sufficiently small, where $\boldsymbol\gamma_k$ is defined as
\begin{equation} \label{eq:change_of_variables}
\boldsymbol\gamma_k = \mathbf{b}_{k+1} - \mathbf{b}_k.
\end{equation}
Applying the approximation in \eqref{eq:function_g_approximation} and the change of variables in \eqref{eq:change_of_variables}, we obtain the Gauss-Newton method:
\begin{equation}
\min_{\mathbf{b}_{k+1}} \left\|  \mathbf{g}(\mathbf{b}_{k+1}) \right\|^2 \approx \min_{\boldsymbol\gamma_k}  \left\|  \mathbf{g}(\mathbf{b}_k) + \frac{d \mathbf{g}}{d \mathbf{b}}(\mathbf{b}_k) \boldsymbol\gamma_k \right\|^2.
\end{equation}
In order to prevent $\boldsymbol\gamma_k$ from being too large, Levenberg \cite{Levenberg1944} and Marquardt \cite{Marquardt1963} propose to add a regularization term (also denoted as damping term) to the cost function, resulting in
\begin{equation} \label{eq:least-squares_problem_damped}
\min_{\boldsymbol\gamma_k} \left\{ \left\|  \mathbf{g}(\mathbf{b}_k) + \frac{d \mathbf{g}}{d \mathbf{b}}(\mathbf{b}_k) \boldsymbol\gamma_k \right\|^2  + \lambda_k \| \boldsymbol\gamma_k \|^2 \right\},
\end{equation}
where $\lambda_k \in \mathbb{R}_{>0}$ is a tuning parameter. Noting that \eqref{eq:least-squares_problem_damped} is a linear least-squares problem, the minimizer of the optimization problem is given by
\begin{equation} \label{eq:least-squares_problem_solution}
\begin{aligned}
\boldsymbol\gamma_k &= - \left( \left( \frac{d \mathbf{g}}{d \mathbf{b}}(\mathbf{b}_k) \right)^T \frac{d \mathbf{g}}{d \mathbf{b}}(\mathbf{b}_k) + \lambda_k \mathbf{I} \right)^{-1} \\
& \quad \times \left( \frac{d \mathbf{g}}{d \mathbf{b}}(\mathbf{b}_k) \right)^T \mathbf{g}(\mathbf{b}_k);
\end{aligned}
\end{equation}
see, for example, \cite[Theorem~2.1.1]{Campbell2009}. By combining \eqref{eq:change_of_variables} and \eqref{eq:least-squares_problem_solution}, we get the update law:
\begin{equation} \label{eq:Levenberg-Marquardt_algorithm}
\begin{aligned}
\mathbf{b}_{k+1} &= \mathbf{b}_k - \left( \left( \frac{d \mathbf{g}}{d \mathbf{b}}(\mathbf{b}_k) \right)^T \frac{d \mathbf{g}}{d \mathbf{b}}(\mathbf{b}_k) + \lambda_k \mathbf{I} \right)^{-1}\\
& \quad \times \left( \frac{d \mathbf{g}}{d \mathbf{b}}(\mathbf{b}_k) \right)^T \mathbf{g}(\mathbf{b}_k).
\end{aligned}
\end{equation}
Hence, the new values $\mathbf{b}_{k+1}$ are equal to the old values $\mathbf{b}_k$ minus the product of a symmetric, positive-definite matrix and the gradient of the cost function in \eqref{eq:least-squares_problem_continuous_rewritten}. Note that the update law in \eqref{eq:Levenberg-Marquardt_algorithm} is similar to that of the Gauss-Newton method for small values of $\lambda_k$. Moreover, the search direction in which the optimization variables are updated is similar to the gradient-descent direction for large values of $\lambda_k$. Due to the uncertainty in the linearized model far from the linearization point, convergence of the algorithm cannot be guaranteed for small values of $\lambda_k$. On the other hand, the decrease in the value of the cost function is small for large values of $\lambda_k$. Therefore, it may take many steps to converge. Various algorithms have been proposed to balance the uncertainty and magnitude of the decrease in cost function value; see \cite{Madsen2004,Pujol2007,Transtrum2012} and references therein. 
\begin{algorithm}[t]
\caption{Standard Levenberg-Marquardt algorithm}
\label{alg:Levenberg-Marquardt_static} 
\begin{algorithmic}[1]
\renewcommand{\algorithmicrequire}{\textbf{Input:}}
\renewcommand{\algorithmicensure}{\textbf{Output:}}
\REQUIRE $\mathbf{b}_0$, $\lambda_0$, (parameters: $\sigma$, $\rho_1$, $\rho_2$)
\ENSURE $\mathbf{b}_k$ $\quad$
\STATE $k\gets 0$
\WHILE {$\left\|2 \left( \frac{d \mathbf{g}}{d \mathbf{b}}(\mathbf{b}_k) \right)^T \mathbf{g}(\mathbf{b}_k)\right\| > \sigma$} 
\STATE \begin{varwidth}[t]{\linewidth}
      $\mathbf{b}_{k+1}' \gets \mathbf{b}_k - \left( \left( \frac{d \mathbf{g}}{d \mathbf{b}}(\mathbf{b}_k) \right)^T \frac{d \mathbf{g}}{d \mathbf{b}}(\mathbf{b}_k) + \lambda_k \mathbf{I} \right)^{-1}$ \par
        \hskip\algorithmicindent $ \times \left( \frac{d \mathbf{g}}{d \mathbf{b}}(\mathbf{b}_k) \right)^T \mathbf{g}(\mathbf{b}_k)$
      \end{varwidth}
\IF {$\|\mathbf{g}(\mathbf{b}_{k+1}')\|^2 < \|\mathbf{g}(\mathbf{b}_k)\|^2$}
\STATE $\mathbf{b}_{k+1} \gets \mathbf{b}_{k+1}'$
\STATE $\lambda_{k+1} \gets \frac{\lambda_k}{\rho_1}$
\ELSE
\STATE $\mathbf{b}_{k+1} \gets \mathbf{b}_{k}$
\STATE $\lambda_{k+1} \gets \rho_2 \lambda_k$
\ENDIF
\STATE $k \gets k + 1$
\ENDWHILE 
\end{algorithmic}
\end{algorithm}
Algorithm~\ref{alg:Levenberg-Marquardt_static} is a simple illustration of an adaptation method for $\lambda_k$. While the gradient of the cost function is larger than a small constant $\sigma > 0$, new values of the optimization variables are generated in Line~3 in accordance with \eqref{eq:Levenberg-Marquardt_algorithm}. If these new values lead to a decrease in cost-function value, the values are accepted and $\lambda_k$ is decreased by a factor $\frac{1}{\rho_1}$. If not, the values are rejected and $\lambda_k$ is increased by a factor $\rho_2$, where $\rho_2 \geq \rho_1 > 1$. Note that robustness measures, such as a positive lower bound on the decrease of the cost-function value in Line~4 and an upper bound on the maximal number of iterations, should be added for practical applications of the algorithm \cite{Madsen2004}. Moreover, the optimization method by Levenberg and Marquardt only converges to local optima. 

\section{Exploiting the structure of the optimization problem to enable parallelization}
\label{sec:exploiting_structure}

If the measurement times span a large interval (i.e., $\overline{t}_m - \underline{t}_m$ is large) and the maximal discretization step $\Delta t$ is small, the number of discretization points $N_d$ is very large. Therefore, the number of variables of the optimization problem in \eqref{eq:least-squares_problem_damped}, which is equal to the dimension $n_\mathbf{b} = n_\mathbf{a} + n_\mathbf{x} \times N_d$ of the vector $\mathbf{b}$ in \eqref{eq:definition_b}, can be very large. In turn, inverting a very large matrix to compute the update in \eqref{eq:Levenberg-Marquardt_algorithm} can be costly. In this section, we present a method to parallelize the computation of the solution of the optimization problem in \eqref{eq:least-squares_problem_damped} to speed up computations.

Let us define
\begin{equation} \label{eq:definition_beta}
\boldsymbol\beta_k(t_d^{(j)}) = \mathbf{x}_{k+1}(t_d^{(j)}) - \mathbf{x}_k(t_d^{(j)}), \quad \boldsymbol\delta_k = \mathbf{a}_{k+1} - \mathbf{a}_k 
\end{equation}
for all $j \in \mathcal{J}_m$, such that $\boldsymbol\gamma_k$ in \eqref{eq:change_of_variables} is given by
\begin{equation} \label{eq:definition_gamma}
\boldsymbol\gamma_k = \begin{bmatrix}
\boldsymbol\beta_k^T(t_d^{(1)}), \, \boldsymbol\beta_k^T(t_d^{(2)}), \, \dots, \, \boldsymbol\beta_k^T(t_d^{(N_d)}), \, \boldsymbol\delta_k^T
\end{bmatrix}^T.
\end{equation}
It follows that the optimization problem in \eqref{eq:least-squares_problem_damped} can be written as
\begin{equation} \label{eq:rewritten_damped}
\min_{\{ \boldsymbol\beta_k \}, \boldsymbol\delta_k} \left\{ \sum_{j = 1}^{N_d} q_{j,k} + r_k \right\},
\end{equation}
with
\begin{equation} \label{eq:qj}
\begin{aligned}
q_{j,k} &= \biggl\| \mathbf{p}_{\mathbf{x},j} + \frac{\partial \mathbf{p}_{\mathbf{x},j}}{\partial \mathbf{x}} \boldsymbol\beta_k(t_d^{(j)}) + \frac{\partial \mathbf{p}_{\mathbf{x},j}}{\partial \mathbf{z}} \boldsymbol\beta_k(t_d^{(j{+}1)})  \\
& \quad  + \frac{\partial \mathbf{p}_{\mathbf{x},j}}{\partial \mathbf{a}} \boldsymbol\delta_k \biggr\|^2  + \lambda_k \| \boldsymbol\beta_k(t_d^{(j)}) \|^2
\end{aligned}
\end{equation}
for all $j \in \mathcal{J}_d\setminus \{N_d\}$, 
\begin{equation} \label{eq:qNd}
\begin{aligned}
q_{N_d,k} &=  \left\| \mathbf{p}_{\mathbf{x},N_d} + \frac{\partial \mathbf{p}_{\mathbf{x},N_d}}{\partial \mathbf{x}} \boldsymbol\beta_k(t_d^{(N_d)}) \right\|^2 + \lambda_k \| \boldsymbol\beta_k(t_d^{(N_d)}) \|^2,
\end{aligned}
\end{equation}
and
\begin{equation} \label{eq:constant_wk}
r_k = \left\| \mathbf{p}_{\mathbf{a}} + \frac{d \mathbf{p}_{\mathbf{a}}}{d \mathbf{a}} \boldsymbol\delta_k \right\|^2 + \lambda_k \| \boldsymbol\delta_k \|^2,
\end{equation}
where $\{ \boldsymbol\beta_k \}$ is short for $\{\boldsymbol\beta_k(t_d^{(j)})\}_{j \in \mathcal{J}_d}$.  
\begin{figure*}[!t]
\addtocounter{equation}{1}
\vspace*{-4pt}
\normalsize
\begin{equation} \label{eq:minimized_partial_sums}
\begin{aligned}
w_{s,k} &= \min_{\{ \boldsymbol\beta_k(t_d^{(j)}) \}_{j = \zeta(s)+1}^{\zeta(s+1)-1}}v_{s,k} = \min_{\{ \boldsymbol\beta_k(t_d^{(j)}) \}_{j = \zeta(s)+1}^{\zeta(s+1)-1}} \Biggl\{ \sum_{j = \zeta(s)}^{\zeta(s+1)-1} q_{j,k} \Biggr\}\\
&= \min_{\boldsymbol\beta_k(t_d^{(\zeta(s+1)-1)})} \Biggl\{  q_{\zeta(s+1)-1,k} + \min_{\boldsymbol\beta_k(t_d^{(\zeta(s+1)-2)})}  \Biggl\{  \dots \, + \min_{\boldsymbol\beta_k(t_d^{(\zeta(s) + 2)})} \Biggl\{  q_{\zeta(s) + 2,k} + \min_{\boldsymbol\beta_k(t_d^{(\zeta(s) + 1)})} \Biggl\{q_{\zeta(s) + 1,k} + q_{\zeta(s),k} \Biggr\}\Biggr\} \dots \Biggr\} \Biggr\}
\end{aligned}
\end{equation}
\vspace*{4pt}
\addtocounter{equation}{2}
\begin{equation} \label{eq:optimizatioN_subproblems}
\begin{aligned}
&\min_{\{ \boldsymbol\beta_k \}, \boldsymbol\delta_k} \left\{ \sum_{j = 1}^{N_d} q_{j,k} + r_k \right\} = \min_{\{ \beta_k(t_d^{(\zeta(s))}) \}_{s = 2}^{N_s}, \boldsymbol\delta_k} \left\{ \sum_{s = 1}^{N_s} w_{s,k} + r_k \right\}\\
&\quad = \min_{\boldsymbol\delta_k} \Biggl\{ r_k + \min_{\boldsymbol\beta_k(t_d^{(\zeta(N_s))})} \Biggl\{  w_{N_s,k} + \min_{\boldsymbol\beta_k(t_d^{(\zeta(N_s{-}1))})} \Biggl\{ \dots  \, + \min_{\boldsymbol\beta_k(t_d^{(\zeta(3))})} \Biggl\{  w_{3,k} + \min_{\boldsymbol\beta_k(t_d^{(\zeta(2))})} \Biggl\{ w_{2,k} + w_{1,k} \Biggr\}  \Biggr\} \dots \Biggr\} \Biggr\} \Biggr\}
\end{aligned}
\end{equation}
\hrulefill
\vspace*{4pt}
\addtocounter{equation}{-5}
\end{figure*}
We have omitted the arguments of the functions $\mathbf{p}_{\mathbf{x},j}$ and $\mathbf{p}_{\mathbf{a}}$ and their derivatives to shorten the expressions in \eqref{eq:rewritten_damped}-\eqref{eq:constant_wk}. Note that the corresponding arguments of $\mathbf{p}_{\mathbf{x},j}$ and its derivatives are $(\mathbf{x}_k(t_d^{(j)}),\mathbf{x}_k(t_d^{(j{+}1)}),\mathbf{a}_k)$ for $j \in \mathcal{J}_d\setminus \{N_d\}$ and $\mathbf{x}_k(t_d^{(N_d)})$ for $j = N_d$, and that the argument of $\mathbf{p}_\mathbf{a}$ and its derivative is $\mathbf{a}_k$. 

Now, to compute the solution of the optimization problem in \eqref{eq:least-squares_problem_damped} using $N_s \in \mathbb{N}_{>0}$ parallel processes, let us divide the sum in \eqref{eq:rewritten_damped} in $N_s$ smaller sums:
\begin{equation} \label{eq:partial_sum_of_terms}
 \sum_{j = 1}^{N_d} q_{j,k} = \sum_{s = 1}^{N_s} v_{s,k}, \quad \mbox{with} \quad v_{s,k} = \sum_{j = \zeta(s)}^{\zeta(s+1)-1} q_{j,k}
\end{equation}
for all $s \in \{1,2,\dots,N_s\}$. Here, $q_{0,k} = 0$, and $\zeta: \mathbb{N} \rightarrow \mathbb{N}$ is a strictly increasing function that satisfies $\zeta(1) = 0$ and $\zeta(N_s+1) = N_d + 1$. The function $\zeta$ is chosen such that the number of summands of each partial sum in \eqref{eq:partial_sum_of_terms} is roughly the same. Instead of minimizing the cost function in \eqref{eq:rewritten_damped} over all optimization variables at once, we minimize the partial sums in \eqref{eq:partial_sum_of_terms} over all optimization variables that do not appear in any of the other partial sums (i.e.,  $\{ \boldsymbol\beta_k(t_d^{(j)}) \}_{j = \zeta(s)+1}^{\zeta(s+1)-1}$ for all $s \in \{1,2,\dots,N_s\}$). For each $s \in \{1,2,\dots,N_s\}$, this leads to the series of nested optimization subproblems with solution $w_{s,k}$ in \eqref{eq:minimized_partial_sums}. 
\addtocounter{equation}{1}
Note that $w_{s,k}$ can be rewritten as
\begin{equation} \label{eq:definition_w}
w_{s,k} = H_{s,\zeta(s{+}1) - \zeta(s),k}
\end{equation}
using the recursion
\begin{equation} \label{eq:subproblem1}
\begin{aligned}
H_{s,j+1,k} = \min_{\boldsymbol\beta_k(t_d^{(\zeta(s)+j)})} \left\{ q_{\zeta(s)+j,k} + H_{s,j,k} \right\}
\end{aligned}
\end{equation}
for all $j \in \{1,2,\dots,\zeta(s{+}1) - \zeta(s)-1\}$, with $H_{s,1,k} = q_{\zeta(s),k}$.
\addtocounter{equation}{1}
Subsequently, the optimization problem in \eqref{eq:rewritten_damped} itself can be formulated as the series of nested optimization subproblems; see \eqref{eq:optimizatioN_subproblems}. Alternatively, we can write \eqref{eq:optimizatioN_subproblems} as
\begin{equation} \label{eq:subproblem2}
\min_{\boldsymbol\delta_k} \left\{ r_k + G_{N_s,k} \right\},
\end{equation}
where $G_{N_s,k}$ is obtained by the recursion
\begin{equation} \label{eq:subproblem3}
G_{s,k} = \min_{\boldsymbol\beta_k(t_d^{(\zeta(s))})} \left\{ w_{s,k} + G_{s-1,k}  \right\}
\end{equation}
for all $s \in\{2,3,\dots,N_s\}$, with $G_{1,k} = w_{1,k}$. We are able to write the optimization problem in \eqref{eq:least-squares_problem_damped} as nested optimization subproblems due to the Markovian structure of the model in \eqref{eq:continuous_state_model}-\eqref{eq:discrete_measurement_model}. The approach of solving an optimization problem by recursively solving optimization subproblems is known as dynamic programming \cite{Bellman1957}. There are two important things to note here. First, all optimization subproblems in \eqref{eq:subproblem1}, \eqref{eq:subproblem2}, and \eqref{eq:subproblem3} are small-scale problems, depending only on a few optimization variables. Therefore, the operative memory required to compute the solution is much lower than for large-scale problem in \eqref{eq:least-squares_problem_damped}. Second, for each $s \in \{1,2,\dots,N_s\}$, the solution $w_{s,k}$ in \eqref{eq:definition_w} can be computed independently and, thus, in parallel. Given sufficient computational power, parallelization greatly reduces the time required to solve the optimization problem. This makes it viable to use much longer measurement series for the identification of the system, which may considerably improve the accuracy of the identified model.

It should be noted that $q_{j,k}$, $r_k$, $w_{s,k}$, $H_{s,j,k}$ and $G_{s,k}$ are all quadratic, sum-of-squares functions of optimization variables. It follows that the minimizers of the optimization subproblems in \eqref{eq:subproblem1}, \eqref{eq:subproblem2} and \eqref{eq:subproblem3} can be computed analytically as linear functions of other optimization variables (i.e., the ones we have not minimized over yet); see Appendix~\ref{app:computation_subproblems}. To be precise, for any $s \in \{1,2,\dots,N_s\}$ and any integer $j$ that satisfies $\zeta(s) < j < \zeta(s+1)$, the minimizer of the subproblem in \eqref{eq:subproblem1} can be written as
\begin{equation} \label{eq:lin_func1}
\boldsymbol\beta_k(t_d^{(j)}) = \left\{ 
\begin{aligned} 
&l_{j,k}(\boldsymbol\beta_k(t_d^{(j+1)}),\boldsymbol\delta_k), & & \mbox{if $s = 1$,}\\
&l_{j,k}(\boldsymbol\beta_k(t_d^{(\zeta(N_s))}),\boldsymbol\delta_k), & & \mbox{if $j = N_d$,}\\
&l_{j,k}(\boldsymbol\beta_k(t_d^{(\zeta(s))}),\boldsymbol\beta_k(t_d^{(j+1)}),\boldsymbol\delta_k), & & \mbox{otherwise}
\end{aligned} \right.
\end{equation}
for some linear function $l_{j,k}$. Similarly, for any $s \in \{2,3,\dots,N_s\}$, the minimizer of the subproblem in \eqref{eq:subproblem3} is given by
\begin{equation} \label{eq:lin_func2}
\boldsymbol\beta_k(t_d^{(\zeta(s))}) = \left\{ 
\begin{aligned} 
&l_{\zeta(s),k}(\boldsymbol\delta_k), & & \mbox{if $s = N_s$,}\\
&l_{\zeta(s),k}(\boldsymbol\beta_k(t_d^{(\zeta(s+1))}),\boldsymbol\delta_k), & & \mbox{otherwise}
\end{aligned} \right.
\end{equation}
for some linear function $l_{\zeta(s),k}$. Because we have minimized the optimization problem over all other variables, the minimizing values of $\boldsymbol\delta_k$ can be directly obtained from \eqref{eq:subproblem2}. We can recursively reconstruct the optimal values of the minimizers $\boldsymbol\beta_k(t_d^{(\zeta(s))})$ in \eqref{eq:subproblem3} by substituting the optimal value for $\boldsymbol\delta_k$ in the linear functions in \eqref{eq:lin_func2}. Subsequently, we can recursively reconstruct the optimal values of the minimizers $\boldsymbol\beta_k(t_d^{(j)})$ in \eqref{eq:subproblem1} by substituting the optimal value for $\boldsymbol\beta_k(t_d^{(\zeta(s))})$ and $\boldsymbol\delta_k$ in the linear functions in \eqref{eq:lin_func1}. Once optimal values of the optimization variables are determined, we may use the equations in \eqref{eq:definition_beta} to update the corresponding state and parameter values for the next iteration.

\begin{algorithm}[t]
\caption{Levenberg-Marquardt algorithm with parallelization}
\label{alg:Levenberg-Marquardt_dynamic} 
\begin{algorithmic}[1]
\renewcommand{\algorithmicrequire}{\textbf{Input:}}
\renewcommand{\algorithmicensure}{\textbf{Output:}}
\REQUIRE $\{\mathbf{x}_0\}$, $\mathbf{a}_0$, $\lambda_0$, (parameters: $\sigma$, $\rho_1$, $\rho_2$)
\ENSURE $\{\mathbf{x}_k\}$, $\mathbf{a}_k$
\STATE $k\gets 0$
\WHILE {$\left\|2 \left( \frac{d \mathbf{g}}{d \mathbf{b}}(\mathbf{b}_k) \right)^T \mathbf{g}(\mathbf{b}_k)\right\| > \sigma$}

\FOR {$s = 1$ to $N_s$}
\STATE $H_{s,1,k} \gets q_{\zeta(s),k}$
\FOR {$j = 1$ to $\zeta(s{+}1)-\zeta(s)-1$}
\STATE Compute $l_{\zeta(s)+j,k}$ in \eqref{eq:lin_func1}
\STATE Compute $H_{s,j+1,k}$ in \eqref{eq:subproblem1}
\ENDFOR
\STATE $w_{s,k} \gets H_{s,\zeta(s{+}1) - \zeta(s),k}$
\ENDFOR
\STATE $G_{1,k} \gets w_{1,k}$
\FOR {$s = 2$ to $N_s$}
\STATE Compute $l_{\zeta(s),k}$ in \eqref{eq:lin_func2}
\STATE Compute $G_{s,k}$ in \eqref{eq:subproblem3}
\ENDFOR
\STATE Compute minimizer $\boldsymbol\delta_k$ in \eqref{eq:subproblem2}
\STATE $\mathbf{a}_{k+1}' \gets \mathbf{a}_k + \boldsymbol\delta_k$
\FOR {$s = N_s$ to $2$}
\STATE Compute $\boldsymbol\beta_k(t_d^{(\zeta(s))})$ by evaluating $l_{\zeta(s),k}$ in \eqref{eq:lin_func2} 
\STATE $\mathbf{x}_{k+1}'(t_d^{(\zeta(s))}) \gets \mathbf{x}_{k}(t_d^{(\zeta(s))}) + \boldsymbol\beta_k(t_d^{(\zeta(s))})$
\ENDFOR
\FOR {$s = 1$ to $N_s$}
\FOR {$j = \zeta(s+1)-1$ to $\zeta(s)+1$}
\STATE Compute $\boldsymbol\beta_k(t_d^{(j)})$ by evaluating $l_{j,k}$ in \eqref{eq:lin_func1}
\STATE $\mathbf{x}_{k+1}'(t_d^{(j)}) \gets \mathbf{x}_{k}(t_d^{(j)}) + \boldsymbol\beta_k(t_d^{(j)})$
\ENDFOR
\ENDFOR

\IF {$\|\mathbf{g}(\mathbf{b}_{k+1}')\|^2 < \|\mathbf{g}(\mathbf{b}_k)\|^2$} 
\STATE $\mathbf{b}_{k+1} \gets \mathbf{b}_{k+1}'$
\STATE $\lambda_{k+1} \gets \frac{\lambda_k}{\rho_1}$
\ELSE
\STATE $\mathbf{b}_{k+1} \gets \mathbf{b}_{k}$
\STATE $\lambda_{k+1} \gets \rho_2 \lambda_k$
\ENDIF
\STATE $k \gets k + 1$
\ENDWHILE 
\end{algorithmic}
\end{algorithm}

An overview of the resulting algorithm is given by Algorithm~\ref{alg:Levenberg-Marquardt_dynamic}. Note that the for loops in Lines~3~to~10 and in Lines~22~to~27 can be parallelized. The same remarks regarding the robustness and convergence of the algorithm apply as for Algorithm~\ref{alg:Levenberg-Marquardt_static} in Section~\ref{sec:solving_problem}.

\section{Network sparsification}
\label{sec:network_spasification}

Consider the system model in \eqref{eq:dynamic_grey-box_model}-\eqref{eq:error_model}, where the error of the first-principle model is modeled by an artificial neural network. To obtain a lean network, we propose to prune the network by sequentially removing network edges. By removing a network edge, we also remove the weight that corresponds to the network edge. Therefore, the number of parameters of the network decreases. Moreover, if all edges of a neuron are removed, then the neuron is no longer part of the network. Hence, removing network edges may also decrease the number of neurons in the network. We aim to only remove the network edges that have little or no effect on the minimal value of the cost function. In that way, we can maintain the quality of the fit after an edge removal. In general, we do not know by how much the minimal value of the cost function changes if any network edges are removed. To find out the difference in minimal value, we need to retrain the network after every change in network structure. Due to the large number of edges and the substantial computational effort it takes to retrain the network, it is often infeasible to retrain the network for every network configuration. In the following section, we present a method to identify which network edges are likely to have the least effect on the minimal value of the cost function. These edges are removed sequentially after verification. This allows us to efficiently identify irrelevant network edges while keeping the overall computational cost relatively low.

\subsection{Estimating the change in minimal cost function value due to a network edge removal}
\label{sec:linearization}

We note that we can effectively remove a network edge by setting its corresponding weight to zero. Because the weights of the network are part of the parameter vector $\mathbf{a}$, determining the minimal value of the cost function after an edge removal is equivalent to determining the minimal cost function value after a certain element of the parameter vector $\mathbf{a}$ is set to zero. Suppose the artificial neural network in \eqref{eq:error_model} is trained using Algorithm~\ref{alg:Levenberg-Marquardt_dynamic} of Section~\ref{sec:exploiting_structure}. Let the optimal state and parameter values be denoted by $\{\mathbf{x}_c\}$ and $\mathbf{a}_c$, or by $\mathbf{b}_c$, for short; see \eqref{eq:definition_b}. Moreover, let the associated cost function value be given by $V_c = \| \mathbf{g}(\mathbf{b}_c) \|^2$; see \eqref{eq:least-squares_problem_continuous_rewritten}. Similar to \eqref{eq:function_g_approximation}, we may approximate the output value of $\mathbf{g}(\mathbf{b})$ at any point $\mathbf{b} = \mathbf{b}_e$ close to $\mathbf{b}_c$ by the first-order Taylor series approximation
\begin{equation} \label{eq:function_g_approximation2}
\mathbf{g}(\mathbf{b}_e) \approx \mathbf{g}(\mathbf{b}_c) + \frac{d \mathbf{g}}{d \mathbf{b}}(\mathbf{b}_c) \boldsymbol\gamma_e
\end{equation}
where $\boldsymbol\gamma_e$ is defined as
\begin{equation} \label{eq:change_of_variables2}
\boldsymbol\gamma_e = \mathbf{b}_e - \mathbf{b}_c.
\end{equation}
For ease of notation, we define
\begin{equation} \label{eq:definition_betae}
\boldsymbol\beta_e(t_d^{(j)}) = \mathbf{x}_e(t_d^{(j)}) - \mathbf{x}_c(t_d^{(j)}), \quad \boldsymbol\delta_e = \mathbf{a}_e - \mathbf{a}_c 
\end{equation}
for all $j \in \mathcal{J}_m$, such that $\boldsymbol\gamma_e$ in \eqref{eq:change_of_variables2} is given by
\begin{equation} \label{eq:definition_gammae}
\boldsymbol\gamma_e = \begin{bmatrix}
\boldsymbol\beta_e^T(t_d^{(1)}), \, \boldsymbol\beta_e^T(t_d^{(2)}), \, \dots, \, \boldsymbol\beta_e^T(t_d^{(N_d)}), \, \boldsymbol\delta_e^T
\end{bmatrix}^T.
\end{equation}
Without loss of generality, we assume that the desired edge is removed from the network by setting the last element of the parameter vector $\mathbf{a} = \mathbf{a}_e$ to zero. Let the last element of $\mathbf{a}_c$ be denoted by $\alpha_c$. If the last element of $\mathbf{a}_e$ is zero and the last element of $\mathbf{a}_c$ is $\alpha_c$, it follows from \eqref{eq:definition_betae} that the last element of $\boldsymbol\delta_e$ is $-\alpha_c$. Equivalently, the last element of $\boldsymbol\gamma_e$ is $-\alpha_c$; see \eqref{eq:definition_gammae}. We let
\begin{equation}
\boldsymbol\delta_e= \begin{bmatrix}
\mathbf{\hat{\boldsymbol\delta}}_e \\ -\alpha_c
\end{bmatrix}, \quad
\boldsymbol\gamma_e = \begin{bmatrix}
\mathbf{\hat{\boldsymbol\gamma}}_e \\ -\alpha_c
\end{bmatrix},
\end{equation}
where $\mathbf{\hat{\boldsymbol\delta}}_e$ and $\mathbf{\hat{\boldsymbol\gamma}}_e$ are vectors containing all remaining elements of $\boldsymbol\delta_e$ and $\boldsymbol\gamma_e$, respectively. Assuming that $\mathbf{b}_e$ is close to $\mathbf{b}_c$ (i.e., assuming that $\boldsymbol\gamma_e$ is small), it follows that the minimal value of the cost function after removing the network edge can be accurately approximated by
\begin{equation} \label{eq:optimization_sparsification}
V_e = \min_{\mathbf{\hat{\boldsymbol\gamma}}_e} \left\| \mathbf{g}(\mathbf{b}_{c}) + \frac{d \mathbf{g}}{d \mathbf{b}}(\mathbf{b}_{c}) \boldsymbol\gamma_e \right\|^2.
\end{equation}
This optimization problem is very similar to the one in \eqref{eq:least-squares_problem_damped}. Thus, a similar approach to computing the solution can be applied.

As in Section~\ref{sec:exploiting_structure}, we may rewrite the optimization problem in \eqref{eq:optimization_sparsification} in the following recursive form:
\begin{equation} \label{eq:subproblem2e}
V_e = \min_{\mathbf{\hat{\boldsymbol\delta}}_e} \left\{ r_e + G_{N_s,e} \right\},
\end{equation}
where $G_{N_s,e}$ is recursively defined as
\begin{equation} \label{eq:subproblem3e}
G_{s,e} = \min_{\boldsymbol\beta_e(t_d^{(\zeta(s))})} \left\{ w_{s,e} + G_{s-1,e}  \right\},
\end{equation}
for all $s \in\{2,3,\dots,N_s\}$, with $G_{1,e} = w_{1,e}$. For $s \in \{1,2,\dots,N_s\}$, the functions $w_{s,e}$ are given by
\begin{equation} \label{eq:definition_we}
w_{s,e} = H_{s,\zeta(s{+}1) - \zeta(s),e},
\end{equation}
where $H_{s,j+1,e}$ is obtained by the recursion
\begin{equation} \label{eq:subproblem1e}
\begin{aligned}
H_{s,j+1,e} = \min_{\boldsymbol\beta_e(t_d^{(\zeta(s)+j)})} \left\{ q_{\zeta(s)+j,e} + H_{s,j,e} \right\}
\end{aligned}
\end{equation}
for all $j \in \{1,2,\dots,\zeta(s{+}1) - \zeta(s)-1\}$, with $H_{s,1,e} = q_{\zeta(s),e}$. Here, for all $j \in \mathcal{J}_d$, the functions $q_{j,e}$ and $r_e$ are similarly defined as the functions $q_{j,k}$ and $r_k$ in \eqref{eq:qj}-\eqref{eq:constant_wk}. The differences are that their arguments are related to the vector $\boldsymbol\gamma_e$ in \eqref{eq:change_of_variables2} instead of the vector $\boldsymbol\gamma_k$ in \eqref{eq:change_of_variables}, that the functions $\mathbf{p}_{\mathbf{x},j}$ and $\mathbf{p}_\mathbf{a}$, that are used in their definitions, have arguments related to $\mathbf{b}_c$ instead of $\mathbf{b}_k$, and that $\lambda_k = 0$.

In addition, we may compute the state and parameter values that correspond to the minimizer of the optimization problem in \eqref{eq:optimization_sparsification}. These can serve as initial conditions for retraining the network if removing the network edge appears beneficial. Therefore, we introduce the linear functions that describe the minimizers of the optimization subproblems in \eqref{eq:subproblem3e} and \eqref{eq:subproblem1e}, which can be used to evaluate the minimizer of the original optimization problem in \eqref{eq:optimization_sparsification}; see Section~\ref{sec:exploiting_structure} for more details. Similar to \eqref{eq:lin_func1},
for any $s \in \{1,2,\dots,N_s\}$ and any integer $j$ that satisfies $\zeta(s) < j < \zeta(s+1)$, the minimizer of the subproblem in \eqref{eq:subproblem1e} is given by
\begin{equation} \label{eq:lin_func1e}
\boldsymbol\beta_e(t_d^{(j)}) = \left\{ 
\begin{aligned} 
&l_{j,e}(\boldsymbol\beta_e(t_d^{(j+1)}),\boldsymbol\delta_e), & & \mbox{if $s = 1$,}\\
&l_{j,e}(\boldsymbol\beta_e(t_d^{(\zeta(N_s))}),\boldsymbol\delta_e), & & \mbox{if $j = N_d$,}\\
&l_{j,e}(\boldsymbol\beta_e(t_d^{(\zeta(s))}),\boldsymbol\beta_e(t_d^{(j+1)}),\boldsymbol\delta_e), & & \mbox{otherwise}
\end{aligned} \right.
\end{equation}
for some linear function $l_{j,e}$. Also, as in \eqref{eq:lin_func2}, for any $s \in \{2,3,\dots,N_s\}$, the minimizer of the subproblem in \eqref{eq:subproblem3} can be written as
\begin{equation} \label{eq:lin_func2e}
\boldsymbol\beta_e(t_d^{(\zeta(s))}) = \left\{ 
\begin{aligned} 
&l_{\zeta(s),e}(\boldsymbol\delta_e), & & \mbox{if $s = N_s$,}\\
&l_{\zeta(s),e}(\boldsymbol\beta_e(t_d^{(\zeta(s+1))}),\boldsymbol\delta_e), & & \mbox{otherwise}
\end{aligned} \right.
\end{equation}
for some linear function $l_{\zeta(s),e}$.

\begin{algorithm}[t]
\caption{Estimating the change in minimal cost function value due to network edge removal}
\label{alg:cost_function_estimate} 
\begin{algorithmic}[1]
\renewcommand{\algorithmicrequire}{\textbf{Input:}}
\renewcommand{\algorithmicensure}{\textbf{Output:}}
\REQUIRE $\{\mathbf{x}_c\}$, $\mathbf{a}_c$
\ENSURE $V_e$, $\{\mathbf{x}_e\}$, $\mathbf{a}_e$

\FOR {$s = 1$ to $N_s$}
\STATE $H_{s,1,e} \gets q_{\zeta(s),e}$
\FOR {$j = 1$ to $\zeta(s{+}1)-\zeta(s)-1$}
\STATE Compute $l_{\zeta(s)+j,e}$ in \eqref{eq:lin_func1e}
\STATE Compute $H_{s,j+1,e}$ in \eqref{eq:subproblem1e}
\ENDFOR
\STATE $w_{s,e} \gets H_{s,\zeta(s{+}1) - \zeta(s),e}$
\ENDFOR
\STATE $G_{1,e} \gets w_{1,e}$
\FOR {$s = 2$ to $N_s$}
\STATE Compute $l_{\zeta(s),e}$ in \eqref{eq:lin_func2e}
\STATE Compute $G_{s,e}$ in \eqref{eq:subproblem3e}
\ENDFOR
\STATE Compute minimizer $\mathbf{\hat{\boldsymbol\delta}}_e$ in \eqref{eq:subproblem2e}
\STATE Compute $V_e$ in \eqref{eq:subproblem2e}
\STATE $\boldsymbol\delta_e \gets \begin{bmatrix}
\mathbf{\hat{\boldsymbol\delta}}_e \\ -\alpha_c
\end{bmatrix}$
\STATE $\mathbf{a}_e \gets \mathbf{a}_c + \boldsymbol\delta_e$
\FOR {$s = N_s$ to $2$}
\STATE Compute $\boldsymbol\beta_e(t_d^{(\zeta(s))})$ by evaluating $l_{\zeta(s),e}$ in \eqref{eq:lin_func2e}
\STATE $\mathbf{x}_e(t_d^{(\zeta(s))}) \gets \mathbf{x}_c(t_d^{(\zeta(s))}) + \boldsymbol\beta_e(t_d^{(\zeta(s))})$
\ENDFOR
\FOR {$s = 1$ to $N_s$}
\FOR {$j = \zeta(s+1)-1$ to $\zeta(s)+1$}
\STATE Compute $\boldsymbol\beta_e(t_d^{(j)})$ by evaluating $l_{j,e}$ in \eqref{eq:lin_func1e}
\STATE $\mathbf{x}_e(t_d^{(j)}) \gets \mathbf{x}_c(t_d^{(j)}) + \boldsymbol\beta_e(t_d^{(j)})$
\ENDFOR
\ENDFOR

\end{algorithmic}
\end{algorithm}

The proposed method for estimating the minimal value of the cost function and the corresponding state and parameter values is summarized in Algorithm~\ref{alg:cost_function_estimate}. Similar to Algorithm~\ref{alg:Levenberg-Marquardt_dynamic}, note that the for loops in Lines~1~to~8 and in Lines~22~to~27 can be parallelized. Moreover, it should be noted that, if we want to compute the minimal cost function value for any other network edge removal instead of the one we have already computed, the results of Lines~1~to~13 of Algorithm~\ref{alg:cost_function_estimate} are identical and may be reused. This makes it relatively fast to compute all possible edge removals. Although Algorithm~\ref{alg:cost_function_estimate} may be applied to predict the minimal cost function value for any possible edge removal, in some cases, it is desirable to consider only a subset of network edges for removal. For example, by prioritizing the removal of edges related to monomials with a high degree, the polynomial output of the network may be of lower degree than without prioritization. In turn, this may improve the extrapolatory properties of the network model outside the measured region of the state. An example of this approach is given in Section~\ref{sec:example_1}.

\subsection{Acceptance criteria for network edge removal}
\label{sec:acceptance_criteria}

After we have estimated which edge removal results in the smallest increase in minimal cost function value, we compute the corresponding true minimal cost function value by retraining the network without the specific edge. We may use an acceptance criterion to decide whether to accept or reject a proposed removal of a network edge. The simplest criterion is solely based on the value of the cost function. A proposed removal of a network edge is accepted only if the value of the cost function after retraining the network with the proposed removal is lower than a preset limit value. Depending on the amount of initial overfitting, one may be able to remove many edges before exceeding the limit value after retraining the network, even if the limit value is only slightly larger than the minimal value of the cost function for the fully connected network. 

Alternatively, we may use an information criterion to assess the quality of the estimate of the model error. In this case, the removal of a certain network edge is accepted only if the quality of the estimate does not decrease. By using an information criterion, both the value of the cost function as well as the number of network parameters are taken into account. There is a large selection of different information criteria to choose from. Several of them are described in detail in \cite{Burnham2002} and \cite{Konishi2008}. The choice of information criterion may significantly influence the sparsification process due to the differences in assumptions on which the various information criteria are derived. A similar approach of quality assessment has been proposed in \cite{Mangan2017}, where Akaike's Information Criterion and the Bayesian Information Criterion are used to evaluate the results of the sparse identification approach in \cite{Brunton2016}.


A third approach of arriving at an acceptance criterion is by cross-validation. Although there are many different forms of cross-validation, the main principle behind cross-validation is that the measurement data is partitioned in a training set, that is solely used for training the neural network, and a validation set (and/or test set; see \cite{Ripley1996}, for example), to evaluate how well the predictions of the trained network generalize to previously unseen data. In that case, a proposed removal of a network edge is only accepted if it results in a lower value of the cost function based on the validation data. Note that determining the cost function value for the validation set requires solving the optimization problem in \eqref{eq:least-squares_continuous_discrete}, where the (fixed) values of the network parameters are determined in the training stage. The use of a validation set avoids the problem of selecting a suitable limit value for the cost function or a suitable information criterion. However, because the training set and validation set must each cover the relevant part of the state space, this may put an unreasonably large demand on the amount of available measurement data.

\section{Examples}
\label{sec:examples}

As discussed in Section~\ref{sec:introduction}, our formulation of the system identification problem in Section~\ref{sec:problem_formulation} and the corresponding solution method in Sections~\ref{sec:exploiting_structure}~and~\ref{sec:network_spasification} allow for significantly lower data requirements and less stringent limitation on the network structure than other methods (e.g., in \cite{Brunton2016} and \cite{Rudy2017}) that are developed to construct a sparse system model. We illustrate our proposed sparse identification approach with two examples that could not have been computed with the methods in \cite{Brunton2016} and \cite{Rudy2017}. 
In Section~\ref{sec:example_1}, we identify the structure of an unknown system without measuring the full state vector, where the identification is based on measurements with different sampling rates. In Section~\ref{sec:example_2}, we train a network to approximate the error in the first-principle model without the requirement that the network model is linear in its parameters. We show that sparsification can simplify a network model while at the same time improve its accuracy by reducing overfitting. 

\subsection{Lorenz system}
\label{sec:example_1}

\begin{figure}[!t]
\centering
\includegraphics[width=8.4cm]{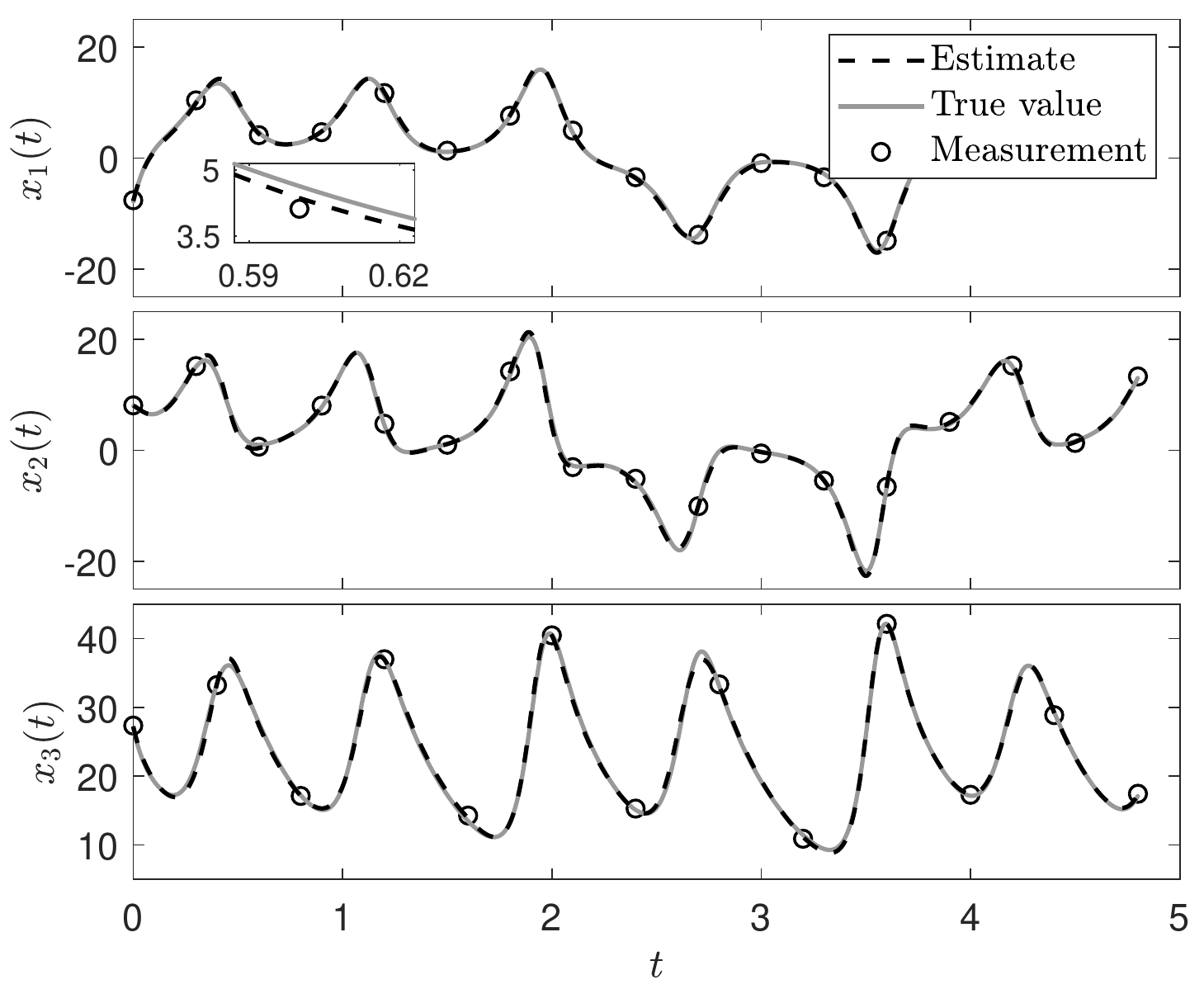}    
\caption{Time signals of the state in \eqref{eq:dynamics_Lorenz} and the corresponding estimates using three state measurements with different sampling rates.}
\label{fig:time_signals_Lorenz_all}
\end{figure}

Consider the Lorenz system:
\begin{equation} \label{eq:dynamics_Lorenz}
\begin{aligned}
\dot{x}_1(t) &= \sigma( x_2(t) - x_1(t) ) \\
\dot{x}_2(t) &=  x_1(t) (\rho - x_3(t) ) - x_2(t) \\
\dot{x}_3(t) &= -\beta x_3(t) + x_1(t) x_2(t),
\end{aligned}
\end{equation}
with state $\mathbf{x}(t) = [x_1(t), \, x_2(t), \, x_3(t)]^T \in \mathbb{R}^3$ and parameters $\sigma = 10$, $\beta = \frac{8}{3}$ and $\rho = 28$. Without any information about the dynamics of the system, the first-principle model assumes a constant state:
\begin{equation}
\mathbf{f}_\text{phys}(\mathbf{x}(t)) = \mathbf{0}.
\end{equation}
This means that the true model error is given by
\begin{equation} \label{eq:error_Lorenz}
\mathbf{e}(t) = \begin{bmatrix}
\sigma( x_2(t) - x_1(t) ) \\
x_1(t) (\rho - x_3(t) ) - x_2(t) \\
-\beta x_3(t) + x_1(t) x_2(t)
\end{bmatrix}.
\end{equation}
We use noisy measurements of the state vector to estimate the model error. The measurements of the first two elements of the state vector are taken every $0.3$ time units from time $t = 0$ to time $t = 4.8$. The measurement of the third element of the state vector is taken every $0.4$ time units on the same time interval. The measurement noise has a Gaussian distribution with zero mean and a variance of one. We use a polynomial network with monomials up to degree two to estimate the model error. A polynomial network is a network with a single hidden layer with monomials as neurons. Each output of the network is a linear combination of monomials. The corresponding monomial gains (i.e., the weights of the edges between the hidden layer and the output layer) are tuned to best match the measurement data. Note that all elements of the error vector \eqref{eq:error_Lorenz} are polynomials of maximal degree two. Therefore, it is possible to obtain an exact expression of the model error using the network. We apply the sparsification approach presented in this paper to identify the model error, with weighting matrices $\mathbf{W}_\mathbf{x} = 10 \mathbf{I}$ and $\mathbf{W}_\mathbf{y} = \mathbf{I}$ and regularization constants $\mu_\mathbf{x} = 10^{-8}$ and $\mu_\mathbf{a} = 10^{-3}$; see Section~\ref{sec:problem_formulation}. The maximal discretization step size is set to $\Delta t = 10^{-3}$.  We use the stage-wise sparsification approach mentioned in Section~\ref{sec:linearization}, where we remove all irrelevant edges connected to monomials with the highest degree in the first stage before we continue to remove irrelevant edges connected to monomials with the second-highest degree in the second stage, etc. Following this approach, we obtain the following network model:
\begin{equation} \label{eq:network_Lorenz_all}
\mathbf{f}_{\text{net}}(\mathbf{x}(t)) = \begin{bmatrix}
-9.378 x_1(t) + 9.483 x_2(t) \\
28.779 x_1(t) - 1.167 x_2(t) - 1.024 x_1(t) x_3(t) \\
- 2.628 x_3(t) + 0.960 x_1(t) x_2(t)
\end{bmatrix}.
\end{equation}
Although we only have a few noisy measurements with a heterogeneous sampling rate at our disposal, the identified network model in \eqref{eq:network_Lorenz_all} has the same structure as the model error in \eqref{eq:error_Lorenz}. Moreover, the identified parameter values are reasonably accurate. The time signals of the state, their estimates and corresponding measurements are depicted in Fig.~\ref{fig:time_signals_Lorenz_all}.

\begin{figure}[!t]
\centering
\includegraphics[width=8.4cm]{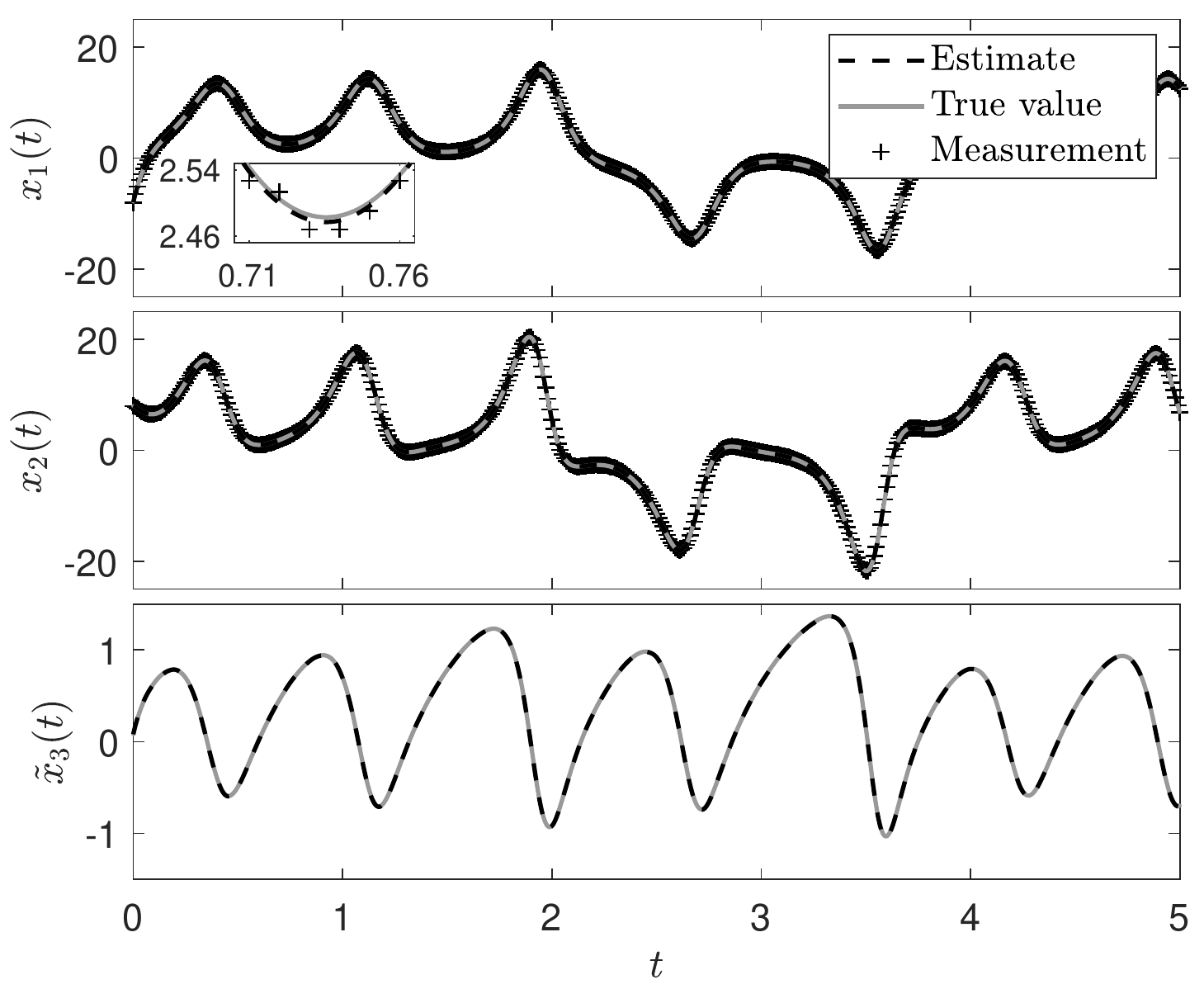}    
\caption{Time signals of the state in \eqref{eq:equivalent_dynamics_Lorenz} and the corresponding estimates using only the first two state measurements.}
\label{fig:time_signals_Lorenz}
\end{figure}

Now, suppose that only measurements of the first two state variables are available (i.e., $\mathbf{y}(t_m)$ $\approx$ $[x_1(t_m), \, x_2(t_m)]^T$). To successfully identify the model error by inferring the dynamics of the third state from the measurements of the first two states, the measurement rate should be increased and the variance of the measurement should be decreased. Let the measurements be taken every $0.01$ time units from time $t = 0$ to time $t = 5$. Moreover, let the noise variance be given by $10^{-4}$. We obtain the following network model:
\begin{equation} \label{eq:network_Lorenz}
\mathbf{f}_{\text{net}}(\mathbf{x}(t)) = \begin{bmatrix}
-9.997 x_1(t) + 9.998 x_2(t) \\
-0.960 x_2(t) + 13.645 x_1(t) x_3(t) \\
5.418 - 2.643 x_3(t) - 0.073 x_1(t) x_2(t)
\end{bmatrix}.
\end{equation}
There seem to be significant differences in structure and value between the error vector in \eqref{eq:error_Lorenz} and the network model in \eqref{eq:network_Lorenz} at first glance. However, using the change of coordinate
\begin{equation}
\tilde{x}_3(t) = \gamma( \rho - x_3(t) ),
\end{equation}
with scaling parameter $\gamma \in \mathbb{R}\setminus\{0\}$, we get the following equivalent form of the dynamics in \eqref{eq:dynamics_Lorenz}:
\begin{equation} \label{eq:equivalent_dynamics_Lorenz}
\begin{aligned}
\dot{x}_1(t) &= \sigma( x_2(t) - x_1(t) ) \\
\dot{x}_2(t) &=  \frac{1}{\gamma}x_1(t)\tilde{x}_3(t) - x_2(t) \\
\dot{\tilde{x}}_3(t) &= \beta ( \gamma \rho  - \tilde{x}_3(t) ) - \gamma x_1(t) x_2(t).
\end{aligned}
\end{equation}
For $\gamma = 0.073$, the corresponding model error is given by
\begin{equation} \label{eq:alternative_error_Lorenz}
\mathbf{\tilde{e}}(t) \approx \begin{bmatrix}
- 10 x_1(t) + 10 x_2(t) \\
- x_2(t) + 13.699 x_1(t) x_3(t) \\
5.451 - 2.667 x_3(t) - 0.073 x_1(t) x_2(t)
\end{bmatrix}.
\end{equation}
Note that the network model in \eqref{eq:network_Lorenz} is an accurate estimate of the error vector in \eqref{eq:alternative_error_Lorenz}. The time signals of the equivalent dynamics in \eqref{eq:equivalent_dynamics_Lorenz} and the corresponding estimates are depicted in Fig.~\ref{fig:time_signals_Lorenz}.
We observe that the differences between the true values of the states and the obtained estimates are small, also for the third state, for which there are no measurements available.

\subsection{Van der Pol oscillator}
\label{sec:example_2}

\begin{figure}[!t]
\centering
\includegraphics[width=3.7cm]{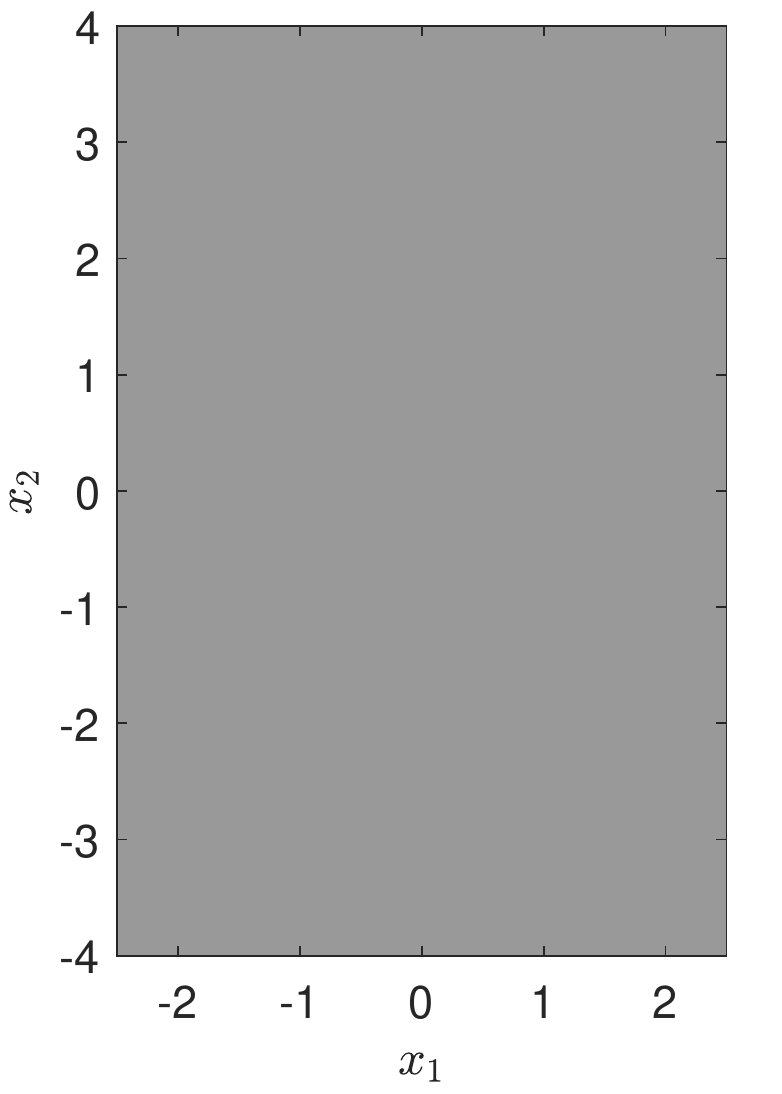}    
~
\includegraphics[width=3.7cm]{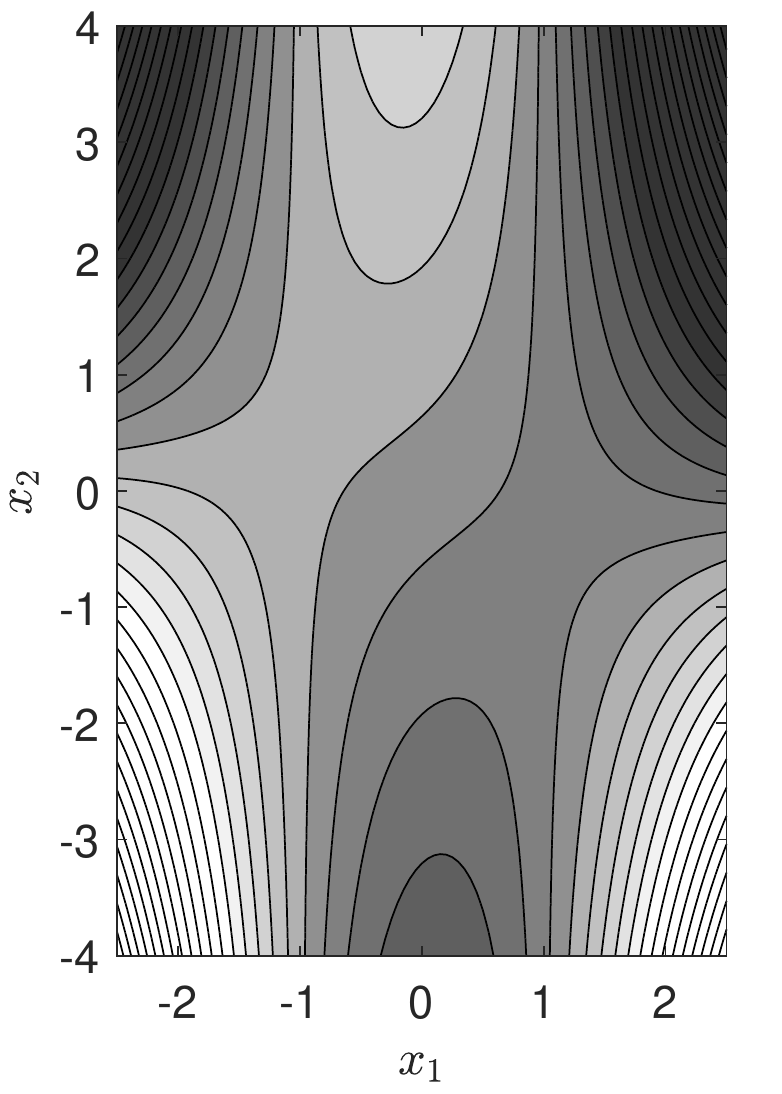} 
\caption{Contour plots of the values of the first (left) and second (right) element of the error vector in \eqref{eq:model_error_Pol}. Note that the first element of the error vector is zero everywhere. Therefore, there are no contours to display. The grey color in the left plot corresponds to a value of zero. Lighter colors in the right plot indicate positive values; darker colors indicate negative values. The values in the right plot range from $-23.5$ to $23.5$.}
\label{fig:fp_error_Pol}
\end{figure}

Consider the forced Van der Pol oscillator:
\begin{equation}
\begin{aligned}
\dot{x}_1(t) &= x_2(t) \\
\dot{x}_1(t) &= A \sin(\omega t) - x_1(t) + \mu (1 - x_1^2(t)) x_2(t),
\end{aligned}
\end{equation}
with state $\mathbf{x}(t) = [x_1(t), \, x_2(t)]^T \in \mathbb{R}^2$ and parameters $A = 1$, $\mu = 1$ and $\omega = 0.2$. Let the first-principle model and the corresponding model error respectively be given by
\begin{equation}
\mathbf{f}_\text{phys}(t,\mathbf{x}(t)) = \begin{bmatrix}
x_2(t) \\
 A \sin(\omega t)
\end{bmatrix}
\end{equation}
and
\begin{equation} \label{eq:model_error_Pol}
\mathbf{e}(t) = \begin{bmatrix}
0 \\
-x_1(t) + \mu (1 - x_1^2(t)) x_2(t)
\end{bmatrix}.
\end{equation}
The error in the first-principle model is visualized in Fig.~\ref{fig:fp_error_Pol}.
It is modeled by an artificial neural network with two hidden layers of 10 units each. The (differentiable) exponential linear unit with shape parameter $a = 1$ (see \cite{Clevert2015}) is used as activation function for all neurons. The inputs of the network are the states $x_1$ and $x_2$. Note that, if the value of $A$ would be uncertain, we could include $\sin(\omega t)$ as an extra input to the network (i.e., $u(t) = \sin(\omega t)$ in \eqref{eq:error_model}) to also learn the model error related to an incorrect estimate of $A$. The outputs of the network are the functions $f_{\text{net},1}(\mathbf{x}(t))$ and $f_{\text{net},2}(\mathbf{x}(t))$ that represent the estimates of the two elements of the error vector. Because there exist no network parameters that represent the model error exactly for all values of $\mathbf{x}$, the network provides only a local approximation of the model error. The network is trained based on noisy measurements of the (full) state of the system. The measurements are taken from time $t = 0$ to time $t = 100$, with a step size of $0.1$ between subsequent measurements. The measurement noise has a Gaussian distribution with a zero mean and a variance of 0.01. 

\begin{figure}[!t]
\centering
\includegraphics[width=8.4cm]{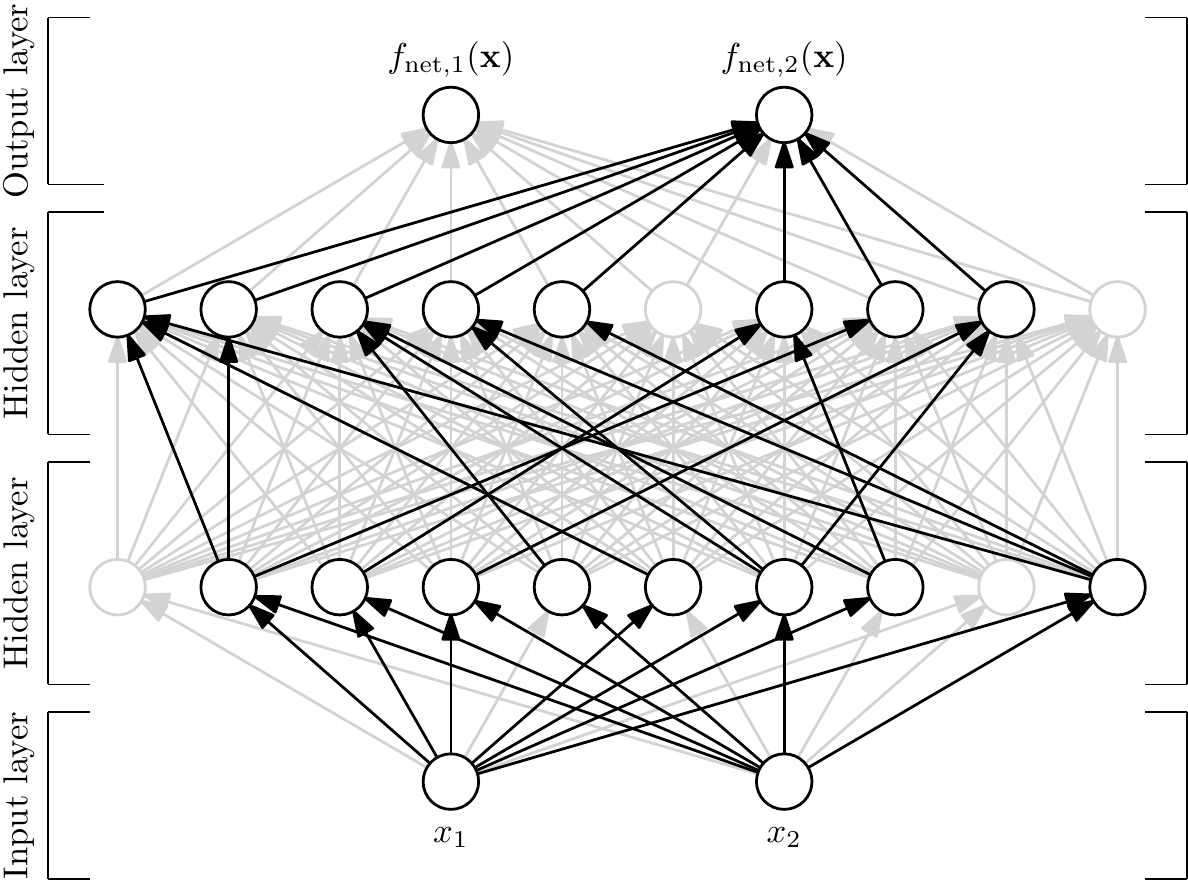}    
\caption{Network layout before sparsification (grey) and after sparsification (black).}
\label{fig:network_layout}
\end{figure}

\begin{figure}[!t]
\centering
\includegraphics[width=4.2cm]{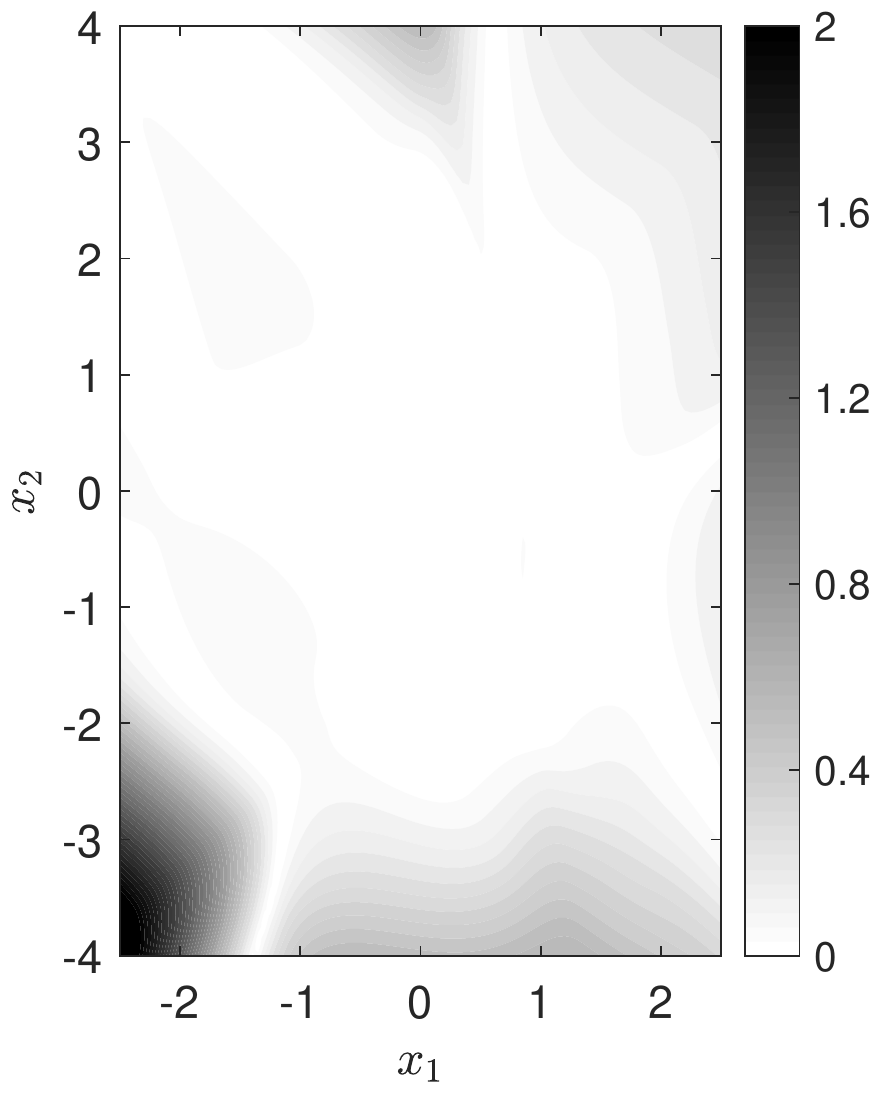}    
~
\includegraphics[width=4.2cm]{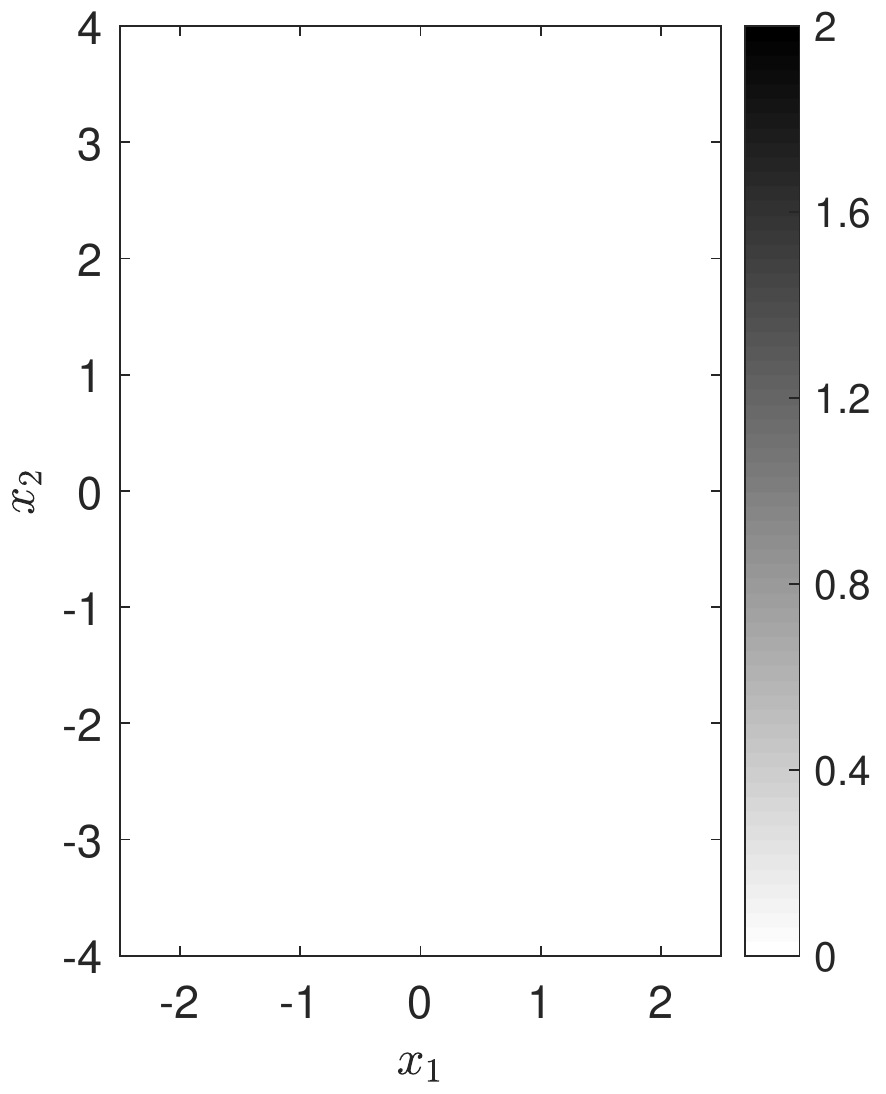} 
\caption{Absolute value of the estimation error for the first element of the error vector in \eqref{eq:model_error_Pol} for the fully connected network (left) and the sparsified network (right). A value of zero is indicated in white. A value of two or higher is indicated in black.}
\label{fig:model_error_Pol_1}
\end{figure}

\begin{figure}[!t]
\centering
\includegraphics[width=4.2cm]{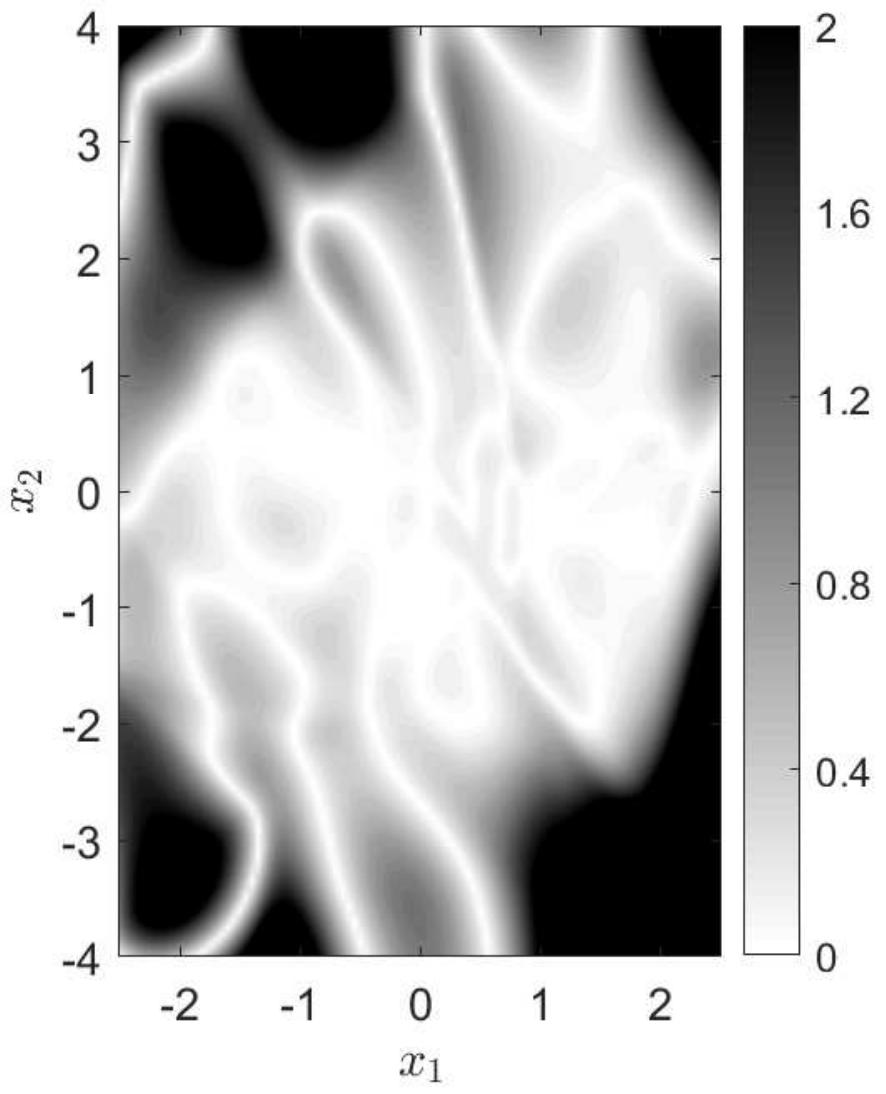}    
~
\includegraphics[width=4.2cm]{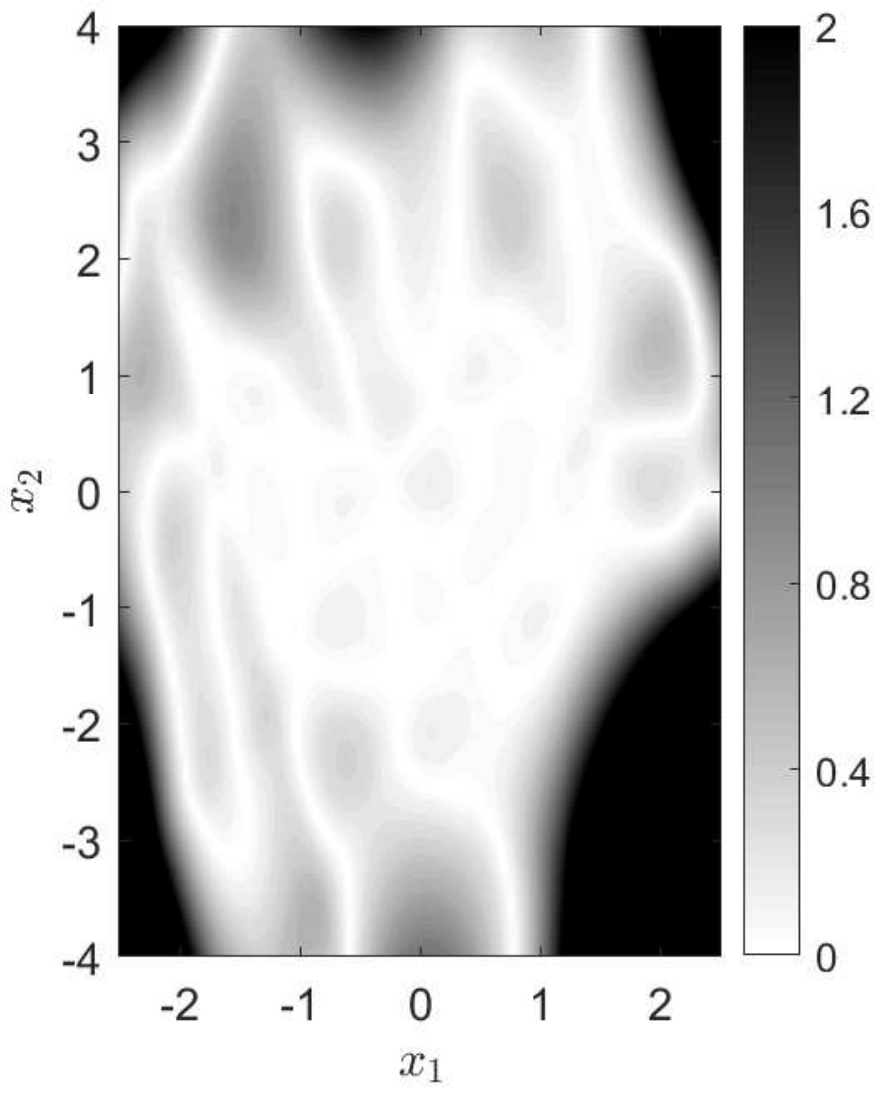} 
\caption{Absolute value of the estimation error for the second element of the error vector in \eqref{eq:model_error_Pol} for the fully connected network (left) and the sparsified network (right). A value of zero is indicated in white. A value of two or higher is indicated in black.}
\label{fig:model_error_Pol_2}
\end{figure}

\begin{figure}[!t]
\centering
\includegraphics[width=3.7cm]{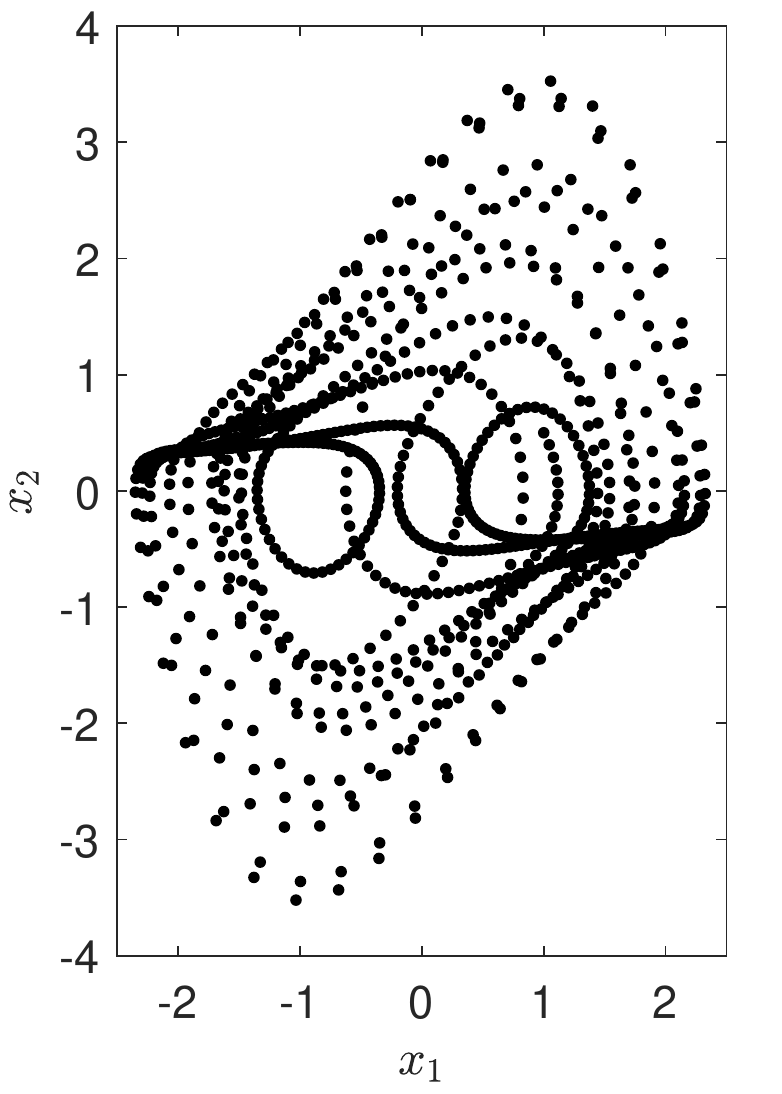}    
\caption{Location of the process measurements in the state space.}
\label{fig:process_measurements}
\end{figure}

By applying the sparsification approach in Section~\ref{sec:network_spasification} with weighting matrices $\mathbf{W}_\mathbf{x} = \mathbf{W}_\mathbf{y} = \mathbf{I}$,  regularization constants $\mu_\mathbf{x} = 0$ and $\mu_\mathbf{a} = 10^{-3}$, and maximal discretization step size $\Delta t = 10^{-3}$, we are able to reduce the number of network edges from 140 to 36 and the number of neurons from 20 to 16. This means that the overall number of network parameters (i.e., weights and biases) is decreased from 160 to 52. The layout of the sparsified network is depicted in Fig.~\ref{fig:network_layout}.
The absolute values of the estimation errors for the two elements of the error vector in \eqref{eq:model_error_Pol} for the fully connected network as well as the sparsified network are depicted in Figs.~\ref{fig:model_error_Pol_1} and \ref{fig:model_error_Pol_2}. 
The state values that correspond to the process measurements are visualized by black dots in Fig.~\ref{fig:process_measurements}. 
Despite having much fewer edges and parameters, we observe that the sparsified network arguably represents the error in the first-principle model better on average in a neighborhood around the process measurements than the fully connected network does. The sparsified network captures the first element of the error vector more accurately than the fully connected network for 100\% of the area of the state space shown in the figures; it captures the second element of the error vector more accurately for roughly 63\% of the area of the state space shown in the figures. This improvement of accuracy can be largely attributed to the reduction of overfitting due to network sparsification.

\section{Discussion}
\label{sec:discussion}

In this paper, we have presented an efficient method for sparse identification of dynamical systems described by ordinary differential equations. We have shown that the Levenberg-Marquardt algorithm, on which our method is based, can be rewritten in a form that enables parallel computation to a large extent. Therefore, the wall time required to solve the identification problem can be greatly decreased. In addition, we have presented a sparsification approach based on backward elimination that utilizes the same time-efficient computation strategy to estimate the relevance of any given network edge, so that only those network edges that have a small effect on the quality of the model fit are removed. We have illustrated with two examples that our method is not affected by many of the limitations, such as fixed sampling rates, full state measurements, and linearity of the model, that plague other methods (e.g., in \cite{Brunton2016} and \cite{Rudy2017}). This flexibility is essential for the applicability of the method in many industrial settings. The main drawback of our approach is the relatively large computational demand. However, utilizing the time-efficient algorithm presented in this paper, the wide applicability of our method is likely to outweigh the larger computational cost in many practical scenarios.

\appendices
\section{Algorithm for generating discretization points}
\label{app:discretization_time}

 Note that all continuous-time signals in \eqref{eq:least-squares_continuous_discrete} are defined on the interval $[\underline{t}_m, \, \overline{t}_m]$. For each subinterval $[t_m^{(i)}, \, t_m^{(i{-}1)}]$, we define the number of discretization points between subsequent measurements
\begin{equation} 
N_r^{(i)} = \left\lceil \frac{t_m^{(i{+}1)} - t_m^{(i)}}{\Delta t} \right\rceil
\end{equation}
and the set of positive integers
\begin{equation}
\mathcal{J}_r^{(i)} = \left\{j \in\mathbb{N}: \, 1 \leq j \leq N_r^{(i)} \right\}
\end{equation}
for all $i \in \mathcal{I}_m \setminus\{N_m\}$. With these definitions in place, let the discretization points between subsequent measurements be given by
\begin{equation} 
\begin{aligned}
\mathcal{T}_r^{(i)} &= \Biggl\{ t_d \in \mathbb{R} : \\
& \qquad (\exists j \in \mathcal{J}_r^{(i)})\left[t_d = t_m^{(i)} + \frac{t_m^{(i{+}1)} - t_m^{(i)}}{N_r^{(i)}}(j-1)\right] \Biggr\}
\end{aligned}
\end{equation}
for all $i \in \mathcal{I}_m \setminus\{N_m\}$. Additionally, let $\mathcal{T}_r^{(N_m)} = \{t_m^{(N_m)}\}$. The set of all discretization points is given by
\begin{equation} \label{eq:definition_Td}
\mathcal{T}_d = \bigcup\limits_{i \in \mathcal{I}_m} \mathcal{T}_r^{(i)}.
\end{equation}
It can be noted that $\mathcal{T}_m \subseteq \mathcal{T}_d$. The cardinality of $\mathcal{T}_d$ is given by 
\begin{equation} \label{eq:definition_Nd}
N_d = \sum_{i \in \mathcal{I}_m} N_r^{(i)},
\end{equation}
where $N_r^{(N_m)} = 1$. 

\section{Algorithm for solving the optimization subproblems}
\label{app:computation_subproblems}

Note that the optimization subproblems in \eqref{eq:subproblem1}, \eqref{eq:subproblem2}, and \eqref{eq:subproblem3} can all be written in the form
\begin{equation} \label{eq:problem_example}
\min_{\mathbf{r}} \left\{ g_1(\mathbf{r},\mathbf{s}) + g_2(\mathbf{r},\mathbf{s}) \right\},
\end{equation}
where $g_1$ and $g_2$ are quadratic, sum-of-squares functions of the optimization variables we want to minimize over $\mathbf{r}$, as well as other optimization variables $\mathbf{s}$. Because $g_1$ and $g_2$ are quadratic and sum of squares, they can be written as
\begin{equation}
g_{i}(\mathbf{r},\mathbf{s}) = \left\| \begin{bmatrix}
\mathbf{M}_{\mathbf{r},i} & \mathbf{M}_{\mathbf{s},i} & \mathbf{M}_{1,i}
\end{bmatrix} \begin{bmatrix}
\mathbf{r} \\ \mathbf{s} \\ 1
\end{bmatrix} \right\|^2
\end{equation}
for $i \in \{1,2\}$ and some coefficient matrices $\mathbf{M}_{\mathbf{r},i}$, $\mathbf{M}_{\mathbf{s},i}$, and $\mathbf{M}_{1,i}$. Note that the sum of $g_1$ and $g_2$ can be written as
\begin{equation}
g_{1}(\mathbf{r},\mathbf{s}) + g_{2}(\mathbf{r},\mathbf{s}) =\left\| \begin{bmatrix}
\mathbf{M}_{\mathbf{r}} & \mathbf{M}_{\mathbf{s}} & \mathbf{M}_{1}
\end{bmatrix} \begin{bmatrix}
\mathbf{r} \\ \mathbf{s} \\ 1
\end{bmatrix} \right\|^2,
\end{equation}
with
\begin{equation}
\mathbf{M}_\mathbf{r} = \begin{bmatrix}
\mathbf{M}_{\mathbf{r},1} \\ \mathbf{M}_{\mathbf{r},2}
\end{bmatrix}, \quad \mathbf{M}_\mathbf{s} = \begin{bmatrix}
\mathbf{M}_{\mathbf{s},1} \\ \mathbf{M}_{\mathbf{s},2}
\end{bmatrix}, \quad \mathbf{M}_1 = \begin{bmatrix}
\mathbf{M}_{1,1} \\ \mathbf{M}_{1,2}
\end{bmatrix}.
\end{equation}
Because the cost function $g_1 + g_2$ is quadratic and sum of squares, all values of $\mathbf{r}$ for which the gradient of the cost function with respect to $\mathbf{r}$ is zero are minimizers. That is, $\mathbf{r}$ minimizes the cost function $g_1 + g_2$ if it satisfies
\begin{equation}
2\mathbf{M}_{\mathbf{r}}^T \left( \mathbf{M}_{\mathbf{r}} \mathbf{r} + \mathbf{M}_{\mathbf{s}} \mathbf{s} + \mathbf{M}_{1} \right) = \mathbf{0}.
\end{equation}
Therefore, a minimizer is given by
\begin{equation} \label{eq:minimizer_example}
\mathbf{r} = - \left( \mathbf{M}_{\mathbf{r}}^T \mathbf{M}_{\mathbf{r}} \right)^+ \mathbf{M}_{\mathbf{r}}^T \begin{bmatrix}
\mathbf{M}_{\mathbf{s}} & \mathbf{M}_{1}
\end{bmatrix} \begin{bmatrix}
\mathbf{s} \\ 1
\end{bmatrix}.
\end{equation}
Note that this minimizer is a linear function of the other optimization variables $\mathbf{s}$. This minimizer is unique if $\mathbf{M}_{\mathbf{r}}^T \mathbf{M}_{\mathbf{r}}$ is invertible, in which case we can exchange the pseudoinverse by the regular inverse. The corresponding solution of the optimization problem in \eqref{eq:problem_example} is given by:
\begin{equation} \label{eq:solution_example}
\min_{\mathbf{r}} \left\{ g_1(\mathbf{r},\mathbf{s}) + g_2(\mathbf{r},\mathbf{s}) \right\} = \left\| \mathbf{P}_\mathbf{r}  \begin{bmatrix}
\mathbf{M}_\mathbf{s} & \mathbf{M}_1
\end{bmatrix} \begin{bmatrix}
\mathbf{s} \\ 1
\end{bmatrix} \right\|^2,
\end{equation}
with
\begin{equation}
\mathbf{P}_\mathbf{r} = \mathbf{I} - \mathbf{M}_\mathbf{r} \left( \mathbf{M}_{\mathbf{r}}^T \mathbf{M}_{\mathbf{r}} \right)^+ \mathbf{M}_{\mathbf{r}}^T.
\end{equation}

\subsection{Computation using the QR-decomposition}

The computation of the minimizer in \eqref{eq:minimizer_example} and solution in \eqref{eq:solution_example} of the optimization problem in \eqref{eq:problem_example} can be simplified using the QR-decomposition. The QR decomposition decomposes any real, rectangular matrix $\mathbf{A}$ into an orthogonal matrix $\mathbf{Q}$ and an upper triangular matrix $\mathbf{R}$ (where the rank of $\mathbf{R}$ is equal to its number of nonzero diagonal elements), such that $\mathbf{A} = \mathbf{Q} \mathbf{R}$; see \cite{Hsieh1993}, for example. Note that $\mathbf{A}^T \mathbf{A} = ( \mathbf{Q} \mathbf{R} )^T \mathbf{Q} \mathbf{R} = \mathbf{R}^T \mathbf{R}$, because $\mathbf{Q}$ is orthogonal. Thus, for any partitioned, real, rectangular matrix
\begin{equation}
\mathbf{A} = \begin{bmatrix}
\mathbf{A}_{1} & \mathbf{A}_{2}
\end{bmatrix},
\end{equation}
we have
\begin{equation} \label{eq:A_R_QR}
\begin{aligned}
\mathbf{A}^T \mathbf{A} &= 
\begin{bmatrix}
\mathbf{A}_{1}^T \mathbf{A}_{1} & \mathbf{A}_{1}^T \mathbf{A}_{2} \\
\mathbf{A}_{2}^T \mathbf{A}_{1} & \mathbf{A}_{2}^T \mathbf{A}_{2}
\end{bmatrix} \\
& = 
\begin{bmatrix}
\mathbf{R}_{11}^T \mathbf{R}_{11} & \mathbf{R}_{11}^T \mathbf{R}_{12} \\ \mathbf{R}_{12}^T \mathbf{R}_{11} & \mathbf{R}_{12}^T \mathbf{R}_{12} + \mathbf{R}_{22}^T \mathbf{R}_{22}
\end{bmatrix} =
\mathbf{R}^T \mathbf{R}
\end{aligned}
\end{equation}
for some partitioned, upper triangular matrix
\begin{equation}
\mathbf{R} = \begin{bmatrix}
\mathbf{R}_{11} & \mathbf{R}_{12} \\ \mathbf{0} & \mathbf{R}_{22}
\end{bmatrix},
\end{equation}
where the elements of $\mathbf{R}$ are implicitly given by \eqref{eq:A_R_QR}.
It follows that
\begin{equation} \label{eq:QR_Z}
\mathbf{R}_{11}^{+} \mathbf{R}_{12} = \left( \mathbf{R}_{11}^T \mathbf{R}_{11} \right)^{+} \mathbf{R}_{11}^T \mathbf{R}_{12} = \left( \mathbf{A}_{1}^T \mathbf{A}_{1} \right)^{+} \mathbf{A}_{1}^T \mathbf{A}_{2}
\end{equation}
and
\begin{equation} \label{eq:QR_S}
\begin{aligned}
\mathbf{R}_{22}^T \mathbf{R}_{22} &= \mathbf{A}_{2}^T \mathbf{A}_{2} - \mathbf{R}_{12}^T \mathbf{R}_{12} \\
& = \mathbf{A}_{2}^T \mathbf{A}_{2} - \mathbf{R}_{12}^T \mathbf{R}_{11} \left( \mathbf{R}_{11}^T \mathbf{R}_{11} \right)^{+} \mathbf{R}_{11}^T \mathbf{R}_{12} \\
& = \mathbf{A}_{2}^T \left( \mathbf{I} -  \mathbf{A}_{1} \left( \mathbf{A}_{1}^T \mathbf{A}_{1} \right)^{+} \mathbf{A}_{1}^T \right) \mathbf{A}_{2}.
\end{aligned}
\end{equation}
Now, if $\mathbf{A} = \begin{bmatrix}
\mathbf{M}_{\mathbf{r}} & \mathbf{M}_{\mathbf{s}} & \mathbf{M}_{1}
\end{bmatrix}$, with $\mathbf{A}_1 = \mathbf{M}_{\mathbf{r}}$ and $\mathbf{A}_2 = \begin{bmatrix} \mathbf{M}_{\mathbf{s}} & \mathbf{M}_{1}
\end{bmatrix}$, by comparison with \eqref{eq:QR_Z} and \eqref{eq:QR_S}, we obtain the following simple expressions for the minimizer in \eqref{eq:minimizer_example} and the solution in \eqref{eq:solution_example}:
\begin{equation} 
\mathbf{r} = - \mathbf{R}_{11}^+ \mathbf{R}_{12} \begin{bmatrix}
\mathbf{s} \\ 1
\end{bmatrix}
\end{equation}
and
\begin{equation} 
\min_{\mathbf{r}} \left\{ g_1(\mathbf{r},\mathbf{s}) + g_2(\mathbf{r},\mathbf{s}) \right\} = \left\| \mathbf{R}_{22} \begin{bmatrix}
\mathbf{s} \\ 1
\end{bmatrix} \right\|^2.
\end{equation}
Hence, the minimizer in \eqref{eq:minimizer_example} and the solution in \eqref{eq:solution_example} can be efficiently computed using the QR-decomposition.

\bibliographystyle{plain}        
\bibliography{reference_list}

%

%
%
%




\end{document}